\documentclass[paper,floatfix,preprintnumbers,nofootinbib,superscriptaddress]{revtex4}

\usepackage{graphicx}
\usepackage{amsmath}
\usepackage{amsfonts}
\usepackage{amssymb}
\usepackage{color}
\usepackage{subfigure}
\usepackage{epsfig}
\usepackage{morefloats}
\usepackage{multirow}
\usepackage{graphicx,booktabs}
\usepackage{mathrsfs}
\usepackage{txfonts}
\usepackage{indentfirst}
\usepackage{graphicx,booktabs}

\usepackage{longtable,lscape}

\usepackage[colorlinks, citecolor=blue,anchorcolor=red,menucolor=red, linkcolor=red,filecolor=red,urlcolor=blue,frenchlinks=red]{hyperref}

\newcommand{\vsig}{\mbox{\boldmath$\sigma$\unboldmath}}

\usepackage{comment}

\begin{document}

%\begin{spacing}{2.0}

\title{Production of excited $\Lambda_c^+$ baryons through $\Lambda_b$ hadronic weak decays}

\author{Kai-Lei Wang}\email{wangkaileicz@foxmail.com}
\affiliation{Department of Physics,
Changzhi University, Changzhi, Shanxi 046011, China}
\affiliation{Synergetic Innovation Center for Quantum Effects and
Applications, Hunan Normal University, Changsha 410081,
China}

\author{Ya-Mei Cao}
\affiliation{Department of Physics,
Changzhi University, Changzhi, Shanxi 046011, China}

\author{Hui-Xiao Duan}
\affiliation{Department of Physics,
Changzhi University, Changzhi, Shanxi 046011, China}

\author{Xian-Hui Zhong}\email{zhongxh@hunnu.edu.cn}
\affiliation{Synergetic Innovation Center for Quantum Effects and
Applications, Hunan Normal University, Changsha 410081,
China}
\affiliation{Department of Physics, Hunan Normal University, and Key
Laboratory of Low-Dimensional Quantum Structures and Quantum Control
of Ministry of Education, Changsha 410081, China }

\date{\today}

\begin{abstract}
In this work, to establish a more abundant $\Lambda_c$ baryon spectrum, we discuss the production potentials of the excited $\Lambda_c$ baryons through $\Lambda_b$ hadronic weak decays within a constituent quark model. Based on our successful explanations of the existing experimental
data for the $\Lambda_b \to \Lambda_c(\pi^-, K^{-}, D^{-}, D^{-}_s, D_s^{*-})$ processes, we further calculate the decay rates of
the $\Lambda_b$ baryon into the $1P$-, $1D$-, $2S$-, $3S$-, $2P$-, $1F$-, and $2D$-wave
$\Lambda_c$ excited states. It is found that these $\Lambda_c$ excitations have large production rates in the $\Lambda_b$ hadronic weak decay process associated with $\pi^-$ meson emitting, i.e., $\Lambda_b\to \Lambda_c^{*}\pi^-$, their branching fractions can reach up to $\mathcal{O}(10^{-3})$. To search for the higher $3S$-, $2P$-, $1F$-, and $2D$-wave $\Lambda_c$ states, the $\Lambda_b \to \Lambda_c (3S) \pi^-\to D^{*0}p \pi^-$, and $\Lambda_b\to \Lambda_c(2P,2D,1F) \pi^-\to D^{(*)0}p\pi^-$ decays are worth observing in future experiments.
\end{abstract}

%\pacs{}

\maketitle

\section{Introduction}

Searching for more new hadron states and exploring the hadron structures are key topics in hadron physics.
For the $\Lambda_{c}^+$ spectrum, since the ground state $\Lambda_{c}(2286)^+$ was first observed in 1976 at Fermilab~\cite{Knapp:1976qw},
lots of studies have been carried out in both theory and experiments (for a review, see~\cite{Chen:2022asf,Cheng:2021qpd,Chen:2016spr,Eichmann:2016yit,Cheng:2015iom,Crede:2013kia,
Klempt:2009pi,Meng:2022ozq}).
Seven candidates of $\Lambda_{c}^+$ excitations, $\Lambda_c(2595)^+$, $\Lambda_c(2625)^+$,
$\Lambda_c(2765)^+$, $\Lambda_c(2860)^+$, $\Lambda_c(2880)^+$, $\Lambda_c(2910)^+$, and $\Lambda_c(2940)^+$ were observed in the past three decades~\cite{pdg}.
It is widely accepted that the $\Lambda_c(2595)^+$ and $\Lambda_c(2625)^+$ should correspond to the two $\lambda$-mode $1P$-wave $\Lambda_{c}^+$ baryons with $J^P=1/2^-$ and $3/2^-$ predicted in the quark model~\cite{Capstick:1986ter,Cheng:2006dk,Cheng:2015naa,Chen:2007xf,Chen:2015kpa,
Chen:2016phw,Chen:2017sci,Ebert:2007nw,Ebert:2011kk,Roberts:2007ni,Chen:2009tm,Chen:2014nyo,Chen:2016iyi,Zhong:2007gp,Wang:2017kfr,Yoshida:2015tia,Xie:2025gom,
Zhang:2024afw,Yu:2023bxn,Shah:2016mig,Tawfiq:1998nk,Ivanov:1999bk,Chow:1995nw,Pirjol:1997nh}.
The $\Lambda_c(2860)^+$ and $\Lambda_c(2880)^+$ resonances favor the two $\lambda$-mode $1D$-wave $\Lambda_{c}^+$ baryons with $J^P=3/2^+$ and $5/2^+$ ~\cite{Lin:2021wrb,Kim:2020imk,Kim:2024tbf,Ponkhuha:2024gms,Arifi:2021orx,Cheng:2006dk,Cheng:2015naa,Chen:2007xf,Chen:2015kpa,
Chen:2016phw,Chen:2017sci,
Ebert:2007nw,Ebert:2011kk,Chen:2009tm,Chen:2014nyo,Chen:2016iyi,Xie:2025gom,Yu:2023bxn,Garcia-Tecocoatzi:2022zrf,Gong:2021jkb,Zhong:2007gp,Yao:2018jmc,
Chen:2017aqm,Guo:2019ytq}. While the $\Lambda_c(2910)^+$ and $\Lambda_c(2940)^+$ resonances may be assigned as the higher $2P$ $\lambda$-mode excitations with $J^P=1/2^-$ and $3/2^-$~\cite{Cheng:2006dk,Chen:2009tm,Yu:2023bxn,Xie:2025gom,Garcia-Tecocoatzi:2022zrf,
Gong:2021jkb,Lu:2019rtg,Lu:2018utx,Yang:2023fsc,Azizi:2022dpn,
Ebert:2007nw,Weng:2024roa,Lu:2016ctt,Guo:2019ytq,Luo:2019qkm,Gandhi:2019xfw,Yu:2022ymb,Zhang:2022pxc,Liu:2009zg}, although there are other
explanations, such as $D^*N$ molecular states, in the literature~\cite{He:2006is,Garcia-Recio:2008rjt,Dong:2009tg,Dong:2010xv,He:2010zq,Liang:2011zza,Dong:2011ys,Ortega:2012cx,Zhang:2012jk,Ortega:2013fta,Dong:2014ksa,Ortega:2014eoa,Xie:2015zga,Yang:2015eoa,
Zhao:2016zhf,Zhang:2019vqe,Wang:2020dhf,Yan:2022nxp,Xin:2023gkf,Ozdem:2023eyz,Yan:2023ttx,Yue:2024paz,Guo:2025tuz,Wang:2022dmw,Wang:2024sbw}. The quark model classification of $\Lambda_c(2765)^+$ is still undetermined, it may be a radial excitation of $\Lambda_c^+$ as suggested in the literature~\cite{Cheng:2006dk,Li:2025frt,Yu:2022ymb,Weng:2024roa}. Currently, except for better understanding the nature of the observed $\Lambda_{c}^+$ baryons, another important task that we face is searching for the missing higher orbital and radial excitations,
such as the $3S$-, $1F$-, and $2D$-wave states, in the $\Lambda_{c}^+$ baryon spectrum.

The LHCb experiments provide a good opportunity for systematically studying the excited $\Lambda_{c}^+$ baryons via $\Lambda_b$ hadronic weak decays. For example, in 2017, the LHCb Collaboration performed an amplitude analysis of the decay $\Lambda_b\to D^0p\pi^-$~\cite{LHCb:2017jym}.  A new resonance $\Lambda_c(2860)^+$ together with two early observed states $\Lambda_c(2880)^+$ and $\Lambda_c(2940)^+$~\cite{CLEO:2000mbh,Belle:2006xni,BaBar:2006itc} were established in the $D^0p$ invariant mass spectrum. With the accumulation of data sample of $\Lambda_b$ baryon in Run 3, more and more higher $\Lambda_{c}^+$ excitations may
have large potentials to be observed in the $\Lambda_b$ weak decays. To provide a useful reference for forthcoming experiments and deeply understand the weak decay mechanism of bottom baryons, we carry out a systematical study of the hadronic weak decays $\Lambda_b\to \Lambda_c^{(*)} h^-$ ($\Lambda_c^{(*)}=\Lambda_c(1S,2S,3S,1P,2P,1D,2D,1F)$, $h^-=\pi^-, \rho^-, K^{-}, K^{\ast-}, D^{-}, D^{\ast-}, D_s^{-}, D_s^{\ast-}$) within a constituent quark model, which has been developed recently, and successfully applied to study the hadronic weak decays of single charm baryons~\cite{Niu:2020gjw,Wang:2022zja,Niu:2021qcc,Niu:2025isf,Niu:2025lgt}, $\Omega_b$ baryon~\cite{Wang:2024ozz}, and charmed mesons~\cite{Cao:2023csx}. This model is similar to the early model that was adopted to deal with the semileptonic decays of heavy $\Lambda_Q$ and
$\Omega_Q$ baryons~\cite{Pervin:2006ie,Pervin:2005ve}.

The hadronic weak decays of $\Lambda_b$ into the ground $\Lambda_{c}^+$ baryon have been widely discussed in the literature with various methods,
such as light-front quark model~\cite{Chua:2019yqh,Ke:2019smy,Zhu:2018jet,Gutsche:2018utw},  heavy quark
effective theory~\cite{Mannel:1992ti,Giri:1997te}, the factorization approach~\cite{Cheng:1996cs,Fayyazuddin:1998ap},
relativistic three-quark model~\cite{Ivanov:1997ra}, covariant oscillator quark model~\cite{Mohanta:1998iu}, perturbative QCD approach~\cite{Zhang:2022iun}, etc.. However, the studies of the $\Lambda_b$ decaying into the excited $\Lambda_{c}^+$ baryons are quite scarce, only a few studies of
the $\Lambda_b\to \Lambda_c(1P,2P,2S) h^-$ processes are found in the literature~\cite{Chua:2019yqh,Li:2022hcn,Chua:2018lfa}.

This paper is organized as follows. In Sec.~\ref{MODEL}, a detailed review of two-body nonleptonic
weak decays of $\Lambda_b \rightarrow \Lambda_c^{(\ast)}h^-$ in
the constituent quark model. In Sec.~\ref{DISSCUS}, the theoretical numerical results and
discussions  are presented. Finally, a short
summary is given in Sec.~\ref{SUM}.

\section{framework}\label{MODEL}

\subsection{The model}

\begin{figure}[htbp]
\centering
\includegraphics[width=0.48\textwidth]{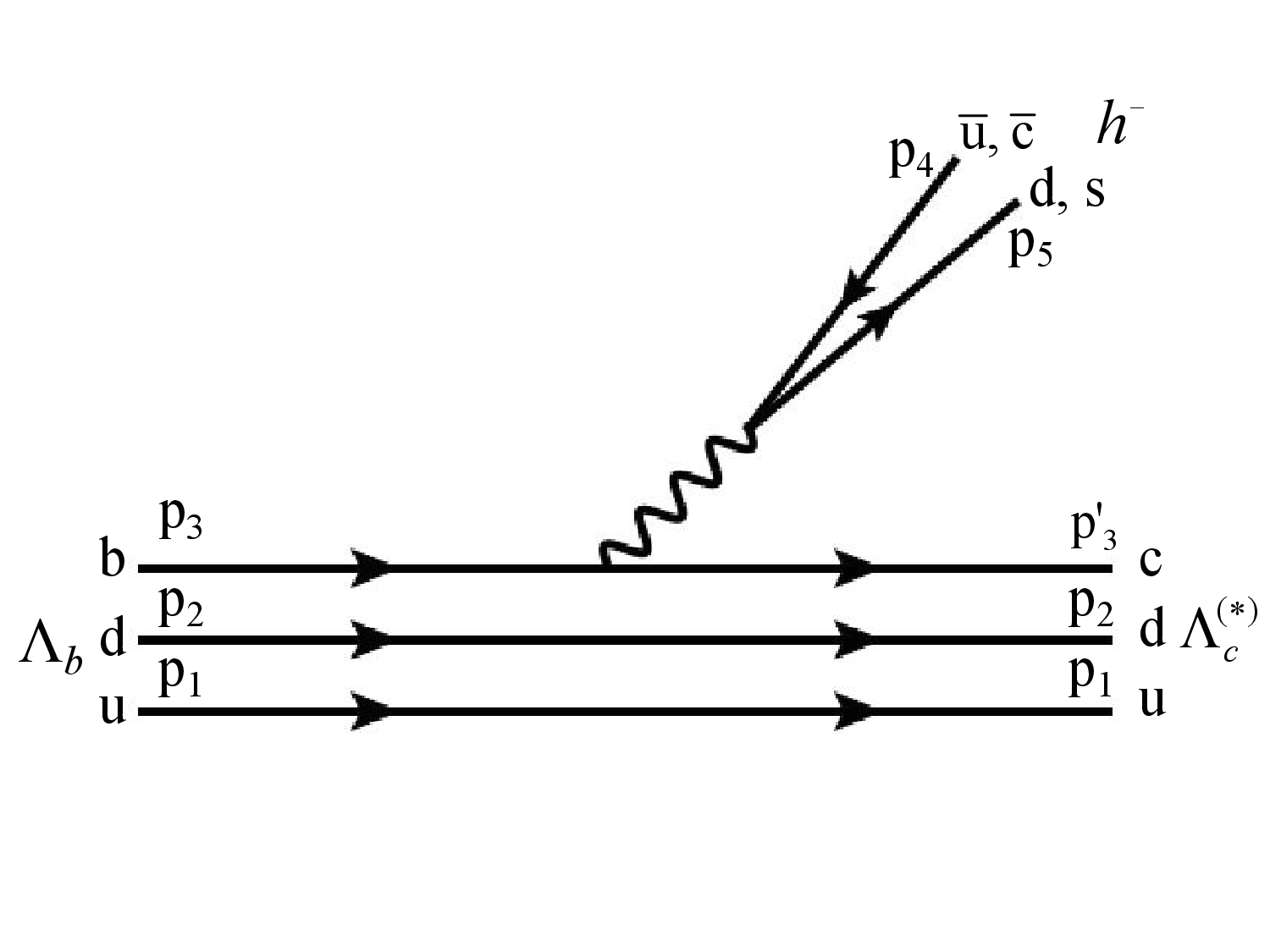}    \vspace{-1.3 cm}
\caption{Feynman diagram for the nonleptonic weak decay processes of $\Lambda_b \rightarrow \Lambda_c^{(\ast)} h^-$  ($h^-=\pi^-$, $\rho^-$, $K^-$, $K^{\ast-}$, $D^-$, $D^{\ast-}$, $D_s^-$, $D_s^{\ast-}$).}\label{tu}
\end{figure}

The $\Lambda_b \to \Lambda_c^{(\ast)}h^-$ decay is particularly interesting because there is only  a $W$-emission diagram contribution~\cite{Cheng:2021qpd}, which is displayed in Fig.~\ref{tu}.
At the quark level, we consider the $b\to c \bar{q}_4q_5~ (\bar{q}_4q_5=\bar{u}d,\bar{u}s,\bar{c}d,\bar{c}s)$ transitions. The effective Hamiltonian is given by~\cite{Buchalla:1995vs}
\begin{eqnarray}\label{dww}
H_W&=&\frac{G_F}{\sqrt{2}}V_{cb} V_{45}^{\ast} (C_1\mathcal{O}_1+C_2\mathcal{O}_2),
\end{eqnarray}
where $G_F$ is the Fermi constant~\cite{pdg},
$C_1$ and $C_2$ are the Wilson coefficients.  $V_{cb}$ and
$V_{45}$ are the Cabibbo-Kobayashi-Maskawa matrix elements~\cite{pdg},
and the current-current operators are
\begin{eqnarray}\label{dww}
\mathcal{O}_1&=&\bar{\psi}_{\bar{q}_{\beta}}\gamma_\mu(1-\gamma_5)\psi_{q_\beta}\bar{\psi}_{\bar{c}_\alpha}\gamma^\mu(1-\gamma_5)\psi_{b_\alpha},\\
\mathcal{O}_2&=&\bar{\psi}_{\bar{q}_\alpha}\gamma_\mu(1-\gamma_5)\psi_{q_\beta}\bar{\psi}_{\bar{c}_\beta}\gamma^\mu(1-\gamma_5)\psi_{b_\alpha},
\end{eqnarray}
with $\psi_{j_\delta}$ ($j=u/d/s/c/b$, $\delta=\alpha/\beta$) representing the $j$-th quark field with a color index $\delta$ in a meson or baryon.

%In the weak decay process of $b$ physics, due to the large $N_c^{eff}$ limit ($N_c^{eff}\to \infty$),  the contribution of $\mathcal{O}_2$ can be negligible. Therefore, we only consider the contribution of $\mathcal{O}_1$.
Considering  its parity behavior, $H_W$  can be divided into two parts, the parity-conserving part $H_{W}^{PC}$ and the parity-violating part $H_{W}^{PV}$~\cite{Niu:2020gjw,Wang:2022zja},
\begin{eqnarray}\label{Hww}
H_{W}=H_{W}^{PC}+H_{W}^{PV}.
\end{eqnarray}
With a nonrelativistic expansion, the two operators can be approximately expressed as
%\begin{eqnarray}\label{Hw}
%H_{W}^{PC}&\simeq&\left(C_1+\frac{C_2}{3}\right)\frac{G_F}{\sqrt{2}}V_{bc}V_{45}^*  \frac{\hat{\phi}_c \hat{O}^f}{(2\pi)^3}\delta^3(\textbf{p}_3-\textbf{p}_3'-\textbf{p}_4-\textbf{p}_5) \Bigg \{\vsig_4 \cdot \left(\frac{\textbf{p}_5}{2m_5}+\frac{\textbf{p}_4}{2m_4}\right)+ \vsig_3 \cdot \left(\frac{\textbf{p}_3'}{2m_3'}+\frac{\textbf{p}_3}{2m_3}\right)\nonumber \\
%& &-\left[\left(\frac{\textbf{p}_3'}{2m_3'}+\frac{\textbf{p}_3}{2m_3}\right)
%-i \vsig_3 \times \left(\frac{\textbf{p}_3}{2m_3}-\frac{\textbf{p}_3'}{2m_3'}\right)\right] \cdot \vsig_4
%-\vsig_3 \cdot \left[\left(\frac{\textbf{p}_5}{2m_5}+\frac{\textbf{p}_4}{2m_4}\right)  -i \vsig_4
%\times \left(\frac{\textbf{p}_4}{2m_4}-\frac{\textbf{p}_5}{2m_5}\right)\right] \Bigg \}\,,\nonumber\\
%%%
%H_{W}^{PV}&\simeq&\frac{G_F}{\sqrt{2}}V_{bc}V_{\bar{q}q}^* \left(C_1+\frac{C_2}{3}\right) \frac{\hat{\phi}_c  \hat{O}^f}{(2\pi)^3}(\vsig_3 \cdot \vsig_4 - 1)\delta^3(\textbf{p}_3-\textbf{p}_3'-\textbf{p}_4-\textbf{p}_5),
%\end{eqnarray}
\begin{eqnarray}\label{Hw}
H_{W}^{PC}&\simeq&\left(C_1+\frac{C_2}{3}\right)\frac{G_F}{\sqrt{2}}V_{cb}V_{45}^*  \frac{\hat{\phi}_c \hat{O}^f}{(2\pi)^3}\delta^3(\textbf{p}_3-\textbf{p}_3'-\textbf{p}_4-\textbf{p}_5) \Bigg \{\vsig_4 \cdot \left(\frac{\textbf{p}_5}{2m_5}+\frac{\textbf{p}_4}{2m_4}-\frac{\textbf{p}_3'}{2m_3'}-\frac{\textbf{p}_3}{2m_3}\right)\nonumber \\
&&+ \vsig_3 \cdot \left(\frac{\textbf{p}_3'}{2m_3'}+\frac{\textbf{p}_3}{2m_3}-\frac{\textbf{p}_5}{2m_5}-\frac{\textbf{p}_4}{2m_4}\right)
+i\vsig_4 \cdot \left[\vsig_3 \times \left(\frac{\textbf{p}_3}{2m_3}-\frac{\textbf{p}_3'}{2m_3'}\right)\right]
+i\vsig_3 \cdot \left[ \vsig_4
\times \left(\frac{\textbf{p}_4}{2m_4}-\frac{\textbf{p}_5}{2m_5}\right)\right] \Bigg \}\,,\nonumber\\
H_{W}^{PV}&\simeq&\frac{G_F}{\sqrt{2}}V_{cb}V_{45}^* \left(C_1+\frac{C_2}{3}\right) \frac{\hat{\phi}_c  \hat{O}^f}{(2\pi)^3}(\vsig_3 \cdot \vsig_4 - 1)\delta^3(\textbf{p}_3-\textbf{p}_3'-\textbf{p}_4-\textbf{p}_5),
\end{eqnarray}
where, $\textbf{p}_j$, $m_j$, and $\vsig_j$ stand for the
momentum, mass, and spin operator of the $j$-th quark, respectively, as shown in
Fig.~\ref{tu}.  While, $\hat{\phi}_c $ and $\hat{O}_f$ stand for the color and flavor operators, respectively. They are
explicitly given by
\begin{eqnarray}\label{OP1}
\hat{O}^f  =  b_5^\dagger(q)b_4^\dagger(\bar{q})b_3^\dagger(c)b_3(b),~\hat{\phi}_c= \delta_{c_4c_5}\delta_{c_{3'}c_3},
\end{eqnarray}
where $\hat{b}^{\dag}$ and $\hat{b}$ are the quark creation and annihilation operators, and the
$\delta_{c_{i}c_j}$ represents that the colors of two quarks, $c_i$ and $c_{j}$, should be conserved.

In the quark model framework, the decay amplitude for the $\Lambda_b \to  \Lambda_c^{(*)}h^-$ is given by
\begin{eqnarray}\label{Hww}
\mathcal{M}^{J_i,J_f,J_{h}}_{M_{J_i},M_{J_f},M_{J_h}}(\Lambda_b \to  \Lambda_c^{(*)}h^-) =  \mathcal{M}^{J_i,J_f,J_{h},PC}_{M_{J_i},M_{J_f},M_{J_h}}+\mathcal{M}^{J_i,J_f,J_{h},PV}_{M_{J_i},M_{J_f},M_{J_h}},
\end{eqnarray}
with $\mathcal{M}^{J_i,J_f,J_{h},PC}_{M_{J_i},M_{J_f},M_{J_h}}=\left\langle \varPsi_{J_fM_{J_f}}^{\Lambda_c^{(*)}} \varPsi_{J_hM_{J_h}}^{h} \left | H_{W}^{PC}\right |\varPsi_{J_iM_{J_i}}^{\Lambda_b} \right\rangle$ and $\mathcal{M}^{J_i,J_f,J_{h},PV}_{M_{J_i},M_{J_f},M_{J_h}}=\left\langle \varPsi_{J_fM_{J_f}}^{\Lambda_c^{(*)}} \varPsi_{J_hM_{J_h}}^{h} \left | H_{W}^{PV} \right |\varPsi_{J_iM_{J_i}}^{\Lambda_b} \right\rangle$,
where $\varPsi_{J_iM_{J_i}}^{\Lambda_b}$, $\varPsi_{J_fM_{J_f}}^{\Lambda_c^{(*)}}$, and $\varPsi_{J_hM_{J_h}}^{h}$ stand for the wave functions of $\Lambda_b$, $\Lambda_c^{(*)}$, and the emitting meson $h^-$, respectively.

The total wave function of a hadron should include
four parts: a color wave function $\zeta_c$, a flavor wave function $\phi_f$, a spin wave function $\chi$, and a spatial wave function $\psi$. The color wave functions for the meson and baryon systems should be a color singlet, which are given by
\begin{equation} \zeta_c=\begin{cases}
        \frac{1}{\sqrt{3}}(R\bar{R}+G\bar{G}+B\bar{B}),   &$for$~\mathrm{meson},\\
        \frac{1}{\sqrt{6}}(RGB-RBG+GBR-GRB+BRG-BGR),   &$for$~\mathrm{baryon},
       \end{cases}
\end{equation}
For the $\Lambda^{(\ast)}_{Q}$ ($Q=c,b$) baryons, the light quarks in which satisfy the SU(3) symmetry,
their flavor wave function belong the $\bar{3}_F$ representation, i.e.,
\begin{eqnarray}
\phi^{\Lambda_Q}_f= \frac{1}{\sqrt{2}}(ud-du)Q .
\end{eqnarray}
For the final meson states ($\pi^-$, $\rho^-$, $K^-$, $K^{*-}$, $D^-$, $D^{*-}$, $D_s^-$, $D_s^{*-}$), the flavor functions are given by
\begin{equation} \phi_f^{h^-}=\begin{cases}
        \bar{u}d,   &$for$~\pi^-,\rho^-\\
        \bar{u}s,   &$for$~K^-,K^{*-}\\
        \bar{c}d,   &$for$~D^-,D^{*-}\\
        \bar{c}s,   &$for$~D_s^-,D_s^{*-}\\
       \end{cases}.
\end{equation}

For a single heavy baryon system $q_1 q_2 Q_3$, the relative momenta between quarks
are defined by
\begin{eqnarray}
\mathbf{p}_\rho&=&\frac{\sqrt{2}}{2}(\textbf{p}_1-\textbf{p}_2)\,,\nonumber\\
\mathbf{p}_\lambda&=&
\frac{\sqrt{6}}{2}\frac{m_3(\textbf{p}_1+\textbf{p}_2)-(m_1+m_2)\textbf{p}_3}{m_1+m_2+m_3}.
\end{eqnarray}
In this work, the spatial wave functions for the $\Lambda_Q^{(\ast)}$ states are used a simple harmonic oscillator (SHO) form.
For a $\Lambda_Q^{(\ast)}$ baryon state with quantum numbers $N$, $L$, and $M_L$,
the spatial wave function can be represented as a product of the $\rho$- and $\lambda$-oscillator parts ~\cite{Mitroy:2013eom,Varga:1997xga,Varga:1995dm,Zhong:2024mnt}:
\begin{eqnarray}\label{Swf}
\psi_{NLM_L}(\textbf{p}_\rho,\textbf{p}_\lambda) = [R_{n_\rho l_\rho}(\textbf{p}_\rho)\mathcal{Y}_{l_{\rho}m_{\rho}}(\textbf{p}_\rho) \otimes R_{n_\lambda l_\lambda}(\textbf{p}_\lambda)\mathcal{Y}_{l_{\lambda} m_{\lambda}}(\textbf{p}_\lambda)]_{LM_L}^N,
\end{eqnarray}
where $n_{\rho/\lambda},l_{\rho/\lambda},m_{\rho/\lambda}$ are the principal, orbital, and magnetic quantum numbers for the $\rho$/$\lambda$ oscillators, respectively, while the quantum numbers $N$, $L$, and $M_L$ satisfy $N=2(n_{\rho}+n_{\lambda})+l_{\rho}+l_{\lambda}$, $L=l_\rho+l_\lambda,...,|l_\rho-l_\lambda|$, and $M_L=m_\rho+m_\lambda$. In Eq.~(\ref{Swf}), the function $\mathcal{Y}_{lm}(\textbf{p})\equiv|\textbf{p}|^{l}Y_{lm}(\mathbf{\hat{p}})$ is the $l$th solid harmonic polynomial,
the SHO form of the radial part $R_{n l}(p)$ in momentum space is given by
\begin{eqnarray}\label{wf}
R_{nl}(p)= (i)^l(-1)^n\left[\frac{2n!}{(n+l+1/2)!}\right]^{1/2}\frac{1}{\alpha^{l+3/2}} \mathrm{exp}\left(-\frac{\textbf{p}^2}{2\alpha^2}\right)L_n^{l+1/2}(\textbf{p}^2/\alpha^2).
\end{eqnarray}
There are two harmonic oscillator parameters $\alpha_\rho$ and $\alpha_\lambda$ in the spatial wave function.
Finally, the total wave function of $\Lambda_{Q}^{(\ast)}$ baryons in the spin-orbital ($L$-$S$) coupling scheme for the states with quantum numbers of total angular momentum and its $z$-component, $J$ and $M_J$, is given by~\cite{Zhong:2025oti}
\begin{eqnarray}
\varPsi_{JM_J}^{\Lambda_{Q}^{(\ast)}} = \mathcal{A}\left \{ [\psi_{NLM_L}(\textbf{p}_\rho,\textbf{p}_\lambda) \otimes \chi^{\sigma}_{SM_S}]_{JM_J} \phi_{f}^{\Lambda_Q} \zeta_{c} \right \},
\end{eqnarray}
where $\mathcal{A}$ represents the operator that enforces antisymmetry on the total wave function when the two light quarks are exchanged, while $S$ and $M_S$ are the quantum numbers of the total spin angular momentum $\boldsymbol{S}$ and its $z$-component, respectively. The explicit forms of the spin wave function $\chi^{\sigma}_{SM_S}$ with different permutation symmetries $\sigma(=s, \rho, \lambda)$ can be found
in Refs.~\cite{Xiao:2013xi, Wang:2017kfr}. The detailed total wave functions of the
$\Lambda_c^{(\ast)}$ baryons, up to the $2D$ states, are presented in Table \ref{Totalwavefunction}.

\begin{table*}[htbp]
\begin{center}
\caption{\label{Totalwavefunction} The total wave functions of the $\Lambda_c$ baryons up to the $2D$ states. The Clebsch-Gordan series for the spin and orbital angular-momentum addition $|JM_J\rangle=\Sigma_{M_{L}+M_{S}=M_J} \langle LM_{L},SM_{S}|J M_J\rangle \ ^N\psi_{LM_L}\chi_{SM_S}$ has been omitted. }
\begin{tabular}{lccccccccccccccccccccccccccccccccccccccccccccc}\hline\hline
State                                          ~~~~~~& $n_\rho$ ~~~~~~& $L_\rho$     ~~~~~~&  $n_\lambda$      ~~~~~~&  $L_\lambda$  ~~~~~~&  $N$  ~~~~~~&  $L$  ~~~~~~& $S$   ~~~~~~& $J^P$      ~~~~~~~~~~~~~~& Wave function     \\\hline
$|\Lambda_c ~~1^2S J^P \rangle$                ~~~~~~& 0   ~~~~~~& 0   ~~~~~~&  0    ~~~~~~&  0   ~~~~~~&  0    ~~~~~~&  0    ~~~~~~& $\frac{1}{2}$                 ~~~~~~& $\frac{1}{2}^+$                     ~~~~~~~~~~~~~~&$^{0}\psi^{S}_{00} \chi^{\rho}_{S_z} \phi \zeta_B$ \\
$|\Lambda_c ~~1^2P_{\lambda} J^P \rangle$      ~~~~~~& 0   ~~~~~~& 0   ~~~~~~&  0    ~~~~~~&  1   ~~~~~~&  1    ~~~~~~&  1    ~~~~~~& $\frac{1}{2}$                 ~~~~~~& $\frac{1}{2}^-$, $\frac{3}{2}^-$    ~~~~~~~~~~~~~~&$^{1}\psi^{\lambda}_{1M_L} \chi^{\rho}_{S_z} \phi \zeta_B$ \\
$|\Lambda_c ~~2^2S_{\lambda} J^P \rangle$      ~~~~~~& 0   ~~~~~~& 0   ~~~~~~&  1    ~~~~~~&  0   ~~~~~~&  2    ~~~~~~&  0    ~~~~~~& $\frac{1}{2}$                 ~~~~~~& $\frac{1}{2}^+$                     ~~~~~~~~~~~~~~&$^{2}\psi^{\lambda}_{00} \chi^{\rho}_{S_z} \phi \zeta_B$ \\
$|\Lambda_c ~~1^2D_{\lambda} J^P \rangle$      ~~~~~~& 0   ~~~~~~& 0   ~~~~~~&  0    ~~~~~~&  2   ~~~~~~&  2    ~~~~~~&  2    ~~~~~~& $\frac{1}{2}$                 ~~~~~~& $\frac{3}{2}^+$, $\frac{5}{2}^+$    ~~~~~~~~~~~~~~&$^{2}\psi^{\lambda}_{2M_L} \chi^{\rho}_{S_z} \phi \zeta_B$ \\
$|\Lambda_c ~~2^2P_{\lambda} J^P \rangle$      ~~~~~~& 0   ~~~~~~& 0   ~~~~~~&  1    ~~~~~~&  1   ~~~~~~&  3    ~~~~~~&  1    ~~~~~~& $\frac{1}{2}$                 ~~~~~~& $\frac{1}{2}^-$, $\frac{3}{2}^-$    ~~~~~~~~~~~~~~&$^{3}\psi^{\lambda}_{1M_L} \chi^{\rho}_{S_z} \phi \zeta_B$ \\
$|\Lambda_c ~~1^2F_{\lambda} J^P \rangle$      ~~~~~~& 0   ~~~~~~& 0   ~~~~~~&  0    ~~~~~~&  3   ~~~~~~&  3    ~~~~~~&  3    ~~~~~~& $\frac{1}{2}$                 ~~~~~~& $\frac{5}{2}^-$, $\frac{7}{2}^-$    ~~~~~~~~~~~~~~&$^{3}\psi^{\lambda}_{3M_L} \chi^{\rho}_{S_z} \phi \zeta_B$ \\
$|\Lambda_c ~~3^2S_{\lambda} J^P \rangle$      ~~~~~~& 0   ~~~~~~& 0   ~~~~~~&  2    ~~~~~~&  0   ~~~~~~&  4    ~~~~~~&  0    ~~~~~~& $\frac{1}{2}$                 ~~~~~~& $\frac{1}{2}^+$                     ~~~~~~~~~~~~~~&$^{4}\psi^{\lambda}_{00} \chi^{\rho}_{S_z} \phi \zeta_B$ \\
$|\Lambda_c ~~2^2D_{\lambda} J^P \rangle$      ~~~~~~& 0   ~~~~~~& 0   ~~~~~~&  1    ~~~~~~&  2   ~~~~~~&  4    ~~~~~~&  2    ~~~~~~& $\frac{1}{2}$                 ~~~~~~& $\frac{3}{2}^+$, $\frac{5}{2}^+$    ~~~~~~~~~~~~~~&$^{4}\psi^{\lambda}_{2M_L} \chi^{\rho}_{S_z} \phi \zeta_B$ \\
\hline\hline
\end{tabular}
\end{center}
\end{table*}

The wave functions for the ground pseudoscalar and vector mesons involved in the final states are given by
\begin{eqnarray}
\Psi_{J_hM_{J_h}}^h=\zeta_c^h \phi_{f}^h\chi_{SM_S}\psi(\textbf{p}_4,\textbf{p}_5),
\end{eqnarray}
where the spatial wave function is adopted as the SHO form, i.e.,
\begin{eqnarray}\label{3wf}
\psi(\textbf{p}_4,\textbf{p}_5)=\frac{2^{3/4}}{\pi^{3/4}\beta_{M}^{3/2}}\mathrm{exp}
\left[-\frac{(\textbf{p}_4-\textbf{p}_5)^2}{4\beta^2}\right],
\end{eqnarray}
where $\beta$ is the well-known HO parameter. For a pseudoscalar meson, the spin wave function is given by
\begin{eqnarray}\label{spwf}
\chi_{00}=\frac{1}{\sqrt{2}}(\uparrow\downarrow-\downarrow\uparrow),
\end{eqnarray}
while for a vector meson, the spin wave functions are given by
\begin{eqnarray}\label{spwf}
\chi_{11}=\uparrow\uparrow, ~~\chi_{10}=\frac{1}{\sqrt{2}}(\uparrow\downarrow+\downarrow\uparrow), ~~\chi_{1-1}=\downarrow\downarrow.
\end{eqnarray}

Then, with the decay amplitudes worked out in the constituent quark model,
the partial decay width for the $\Lambda_b \to  \Lambda_c^{(*)}h^-$ process is calculated by
\begin{eqnarray}\label{dww}
\Gamma &=& 8\pi^2\frac{|\textbf{q}|E_{h}E_{\Lambda_c^{(*)}}}{M_{\Lambda_b}}\frac{1}{2J_{\Lambda_b}+1} \sum_{M_{J_i}=M_{J_f}+M_{J_h}}|\mathcal{M}^{J_i,J_f,J_{h}}_{M_{J_i},M_{J_f},M_{J_h}}|^2,
\end{eqnarray}
where $\textbf{q}$ and $E_{h}$ are the three-momentum and energy of the meson $h$, respectively,
$E_{\Lambda_c^{(*)}}$ is the energy of the $\Lambda_c^{(*)}$ baryon,
and $M_{\Lambda_b}$ is the mass of the $\Lambda_b$ baryon.

\subsection{Model parameters}

In this work, the CKM matrix elements and the lifetime of the $\Lambda_b^0$ baryon are taken from the PDG~\cite{pdg}, which are listed
as follows:
\begin{eqnarray}
&&
(V_{ud},V_{us}) = (0.9737,0.2245),\nonumber\\
&&
(V_{cd},V_{cs},V_{cb}) = (0.221,0.987,0.041), \\
&&
\tau_{\Lambda_b^0} = 1.468\times 10^{-12} s.\nonumber
\end{eqnarray}
The Wilson coefficients $(C_1, C_2)= (1.1, -0.25)$ are taken from Refs.~\cite{Gutsche:2018utw,Buchalla:1995vs,Altmannshofer:2008dz}.

In the calculation, the standard quark model parameters
are adopted. The constituent masses for the $u/d$, $s$, $c$, and $b$
are taken as $m_{u/d}=450$ MeV, $m_s=550$ MeV, $m_c=1450$ MeV, and $m_b=4800$ MeV, which are consistent with those
adopted for the recent heavy-light meson spectrum calculations within the potential model~\cite{Ni:2023lvx}.
For the $\Lambda_b$ and $\Lambda_c^{(\ast)}$ baryons, the $\rho$-mode HO parameter is set to the commonly used value in the literature, i.e. $\alpha_{\rho} = 400$ MeV, while
the $\lambda$-mode HO parameter $\alpha_\lambda$ is related to that of $\rho$ mode with the following relation~\cite{Wang:2017kfr,Yao:2018jmc}
\begin{eqnarray}\label{wf}
\alpha_{\lambda}=\sqrt[4]{\frac{3m_{b(c)}}{m_u+m_d+m_{b(c)}}}\alpha_{\rho}.
\end{eqnarray}
For the mesons, the HO parameter $\beta$ is determined by~\cite{Zhong:2008kd}
\begin{eqnarray}\label{wf}
\beta = \sqrt{\frac{2m_qm_{\bar{q}}}{(m_q+m_{\bar{q}})m_u}} \alpha,
\end{eqnarray}
where the parameter $\alpha$ corresponds to the HO parameter of the $\pi^-$ and $\rho^-$ mesons.
To maintain consistency with baryons, $\alpha$ is also taken as $\alpha =\alpha_{\rho}= 400$ MeV.

The masses for the baryons ($\Lambda_{b}^0$ and $\Lambda_{c}^{(*)+}$) and mesons ($\pi^-$, $\rho^-$, $K^-$, $K^{*-}$,
$D^-$, $D^{*-}$, $D_s^-$, $D_s^{*-}$) observed by experiments are taken from the PDG~\cite{pdg}. For the unestablished $\Lambda_c^*$ resonances,
the masses are taken from the quark model predictions~\cite{Ebert:2011kk}. The corresponding numerical values are listed in Tables~\ref{a1} and ~\ref{result}.

It should be pointed out that there are some uncertainties in our predictions due to the uncertainties of the model parameters.
Our analysis shows that the uncertainties of our predictions mainly come from the HO parameters $\alpha_{\rho}$ and $\beta$ adopted in the calculations. Usually, these HO parameters have an uncertainty of  $\sim10\%$.
Taking a $10\%$ variation for the HO parameters $\alpha_{\rho}$ and $\beta$,  we further evaluate the uncertainties
of our predictions, which are given in Tabs.~\ref{result1S} and \ref{result}. It is found that
uncertainties of the branching fractions can reach up to $\sim 50\%$ due to the uncertainties of the HO parameters.
However, the relative branching ratio $\frac{\mathcal{B}[\Lambda_b^0 \to \Lambda_c^{(\ast)+} M]}{\mathcal{B}[\Lambda_b^0 \to \Lambda_c^+ M]}$,
as shown in Tab.~\ref{result}, bears a small uncertainty of $\sim 10\%$ due to the variation of the baryon HO parameter $\alpha_{\rho}$, however,
is insensitive to the meson HO parameter $\beta$.

%It should be pointed out that there are some uncertainties in our predictions due to the uncertainties of the model parameters.
%Our analysis shows that the uncertainties of our predictions mainly come from the HO parameters $\alpha_{\rho}$ and $\beta$ adopted in the calculations. Usually, these HO parameters have an uncertainty of about $\sim10\%$.
%Taking a $\pm10\%$ variation for the HO parameters $\alpha_{\rho}$ and $\beta$, respectively, we further evaluate the uncertainties
%of our predictions, which are given in Tabs.~\ref{result1S} and \ref{result}. It is found that
%uncertainties of the branching fractions can reach up to $\sim 50\%$ due to the uncertainties of the HO parameters.
%However, the relative branching ratio $\frac{\mathcal{B}[\Lambda_b^0 \to \Lambda_c^{(\ast)+} M]}{\mathcal{B}[\Lambda_b^0 \to \Lambda_c^+ M]}$,
%as shown in Tab.~\ref{result}, bears a small uncertainty of $\sim 10\%$ due to the variation of the baryon HO parameter $\alpha_{\rho}$, however,
%is insensitive to the meson HO parameter $\beta$.
%Consequently, some of our results should be regarded as only semiquantitative estimates based on SU(3) flavor symmetry.
%Fortunately, the predicted branching fractions for $\Lambda_b \to \Lambda_c (\pi^-, K^-, D^-, D_s^-, D_s^{-})$ are in agreement with the available experimental data~\cite{pdg} within the present framework, indicating that the constituent quark model still provides a useful approach for studying $\Lambda_b \to \Lambda_c^{(\ast)} M$ transitions.

\begin{table}[htp]
\begin{center}
\caption{\label{a1} The masses (MeV) for the observed baryons and mesons taken from the PDG~\cite{pdg}. }
\begin{tabular}{cccccccccccccccccccccccccc}\hline\hline
State     ~~~&$\Lambda_b^0$     ~~~&$\pi^-$  ~~~&$K^-$  ~~~&$D^-$   ~~~&$D_s^-$  ~~~&$\rho^-$   ~~~&$K^{\ast-}$   ~~~&$D^{\ast-}$   ~~~&$D_s^{\ast-}$ \\\hline
Mass ~~~&5620       ~~~&140      ~~~&494     ~~~&1870     ~~~&1968     ~~~&775      ~~~&896      ~~~&2007     ~~~&2112      \\
\hline\hline
State     ~~~& $\Lambda_c^+$ ~~~&$\Lambda_c(2595)^+$  ~~~&$\Lambda_c(2625)^+$   ~~~&$\Lambda_c(2765)^+$  ~~~&$\Lambda_c(2860)^+$  ~~~&$\Lambda_c(2880)^+$   ~~~&$\Lambda_c(2910)^+$  ~~~&$\Lambda_c(2940)^+$       \\\hline
Mass ~~~&2286  ~~~&2592       ~~~&2628  ~~~&2767      ~~~&2856     ~~~&2882     ~~~&2914     ~~~&2940                \\
\hline\hline
\end{tabular}
\end{center}
\end{table}

\section{Numerical results and discussion}\label{DISSCUS}

\begin{table*}[htbp]
\begin{center}
\caption{\label{result1S}  Branching fractions (with a unit of $10^{-3}$) of the $\Lambda_b\rightarrow \Lambda_c h^-$ processes ($h^-=\pi^-$, $\rho^-$, $K^-$, $K^{*-}$, $D^-$, $D^{*-}$, $D_s^-$, $D_s^{*-}$). The first error of our predictions comes from the baryon HO parameter $\alpha_{\rho}$ with uncertainties of $\pm 10\%$. The second error of our predictions comes from the meson HO parameter $\beta$
with uncertainties of $\pm 10\%$. For a comparison, the other model results and experimental data are included.}
\begin{tabular}{lccccccccccccccccccccccccccccccccccccccccccccc}\hline\hline
Decays      & This work   &~~~\cite{Chua:2019yqh}   & ~~~\cite{Mannel:1992ti}
          & ~~~\cite{Cheng:1996cs}
          & ~~~\cite{Ivanov:1997ra}
          & ~~~\cite{Giri:1997te}
          & ~~~\cite{Fayyazuddin:1998ap}
          & ~~~\cite{Mohanta:1998iu}
          & ~~~\cite{Zhu:2018jet}
          & ~~~\cite{Gutsche:2018utw}
          & ~~~\cite{Ke:2019smy}    &~~~Exp~\cite{pdg}.   \\ \hline
$\Lambda_b \rightarrow \Lambda_c \pi^-$                  &$5.67^{+3.17+1.88}_{-2.55-1.55}$
          & $4.16_{-1.73}^{+2.43}$
          & $4.6^{+2.0}_{-3.1}$
          & $4.6$
          & $5.62$
          & 3.91
          &
          & $1.75$
          & $4.96$
          &
          & 5.67      &$4.9 \pm 0.4 $        \\
$\Lambda_b \rightarrow \Lambda_c K^-$                    &$0.31^{+0.17+0.10}_{-0.14-0.08}$
          & $0.31_{-0.13}^{+0.18}$
         &&&&&
          & $0.13$
          & $0.393$
          &
          & 0.46      &$0.356 \pm 0.028$         \\
$\Lambda_b \rightarrow \Lambda_c D^-$                    &$0.43^{+0.14+0.14}_{-0.14-0.12}$
          & $0.47_{-0.21}^{+0.30} $ %& $0.53_{-0.22}^{+0.32}$
         &&&&&
          & $0.30$
          & $0.522$
          &
          & 0.76          &$0.46 \pm 0.06$         \\
$\Lambda_b \rightarrow \Lambda_c D_s^-$                  &$10.3^{+3.10+3.39}_{-3.08-2.78}$
          & $11.92_{-5.28}^{+7.69} $  %& $13.50_{-5.70}^{+8.08}$
          & $23^{+3}_{-4}$
          & $13.7$
          &
          & $12.91$
          & $22.3$
          & 7.70
          & $12.4$
          & 14.78
          & 19.94       &$11.0 \pm 1.0 $         \\
$\Lambda_b \rightarrow \Lambda_c D_s^{\ast-}$             &$17.5^{+4.91+5.80}_{-4.97-4.74}$
          & $ 17.49_{-7.48}^{+10.60} $ %& $ 18.62_{-7.81}^{+10.95} $
          & $17.3^{+2.0}_{-3.0}$
          & $21.8$
          &
          & $19.83$
          & $32.6$
          & 14.14
          & $10.5$
          & $25.16$
          & 30.86      & $ 18.3\pm 1.8 $        \\
$\Lambda_b \rightarrow \Lambda_c \rho^-$                 &$6.91^{+3.08+1.62}_{-3.39-2.24}$
          & $ 12.28_{-5.11}^{+7.19} $
          & $6.6^{+2.4}_{-4.0}$
          & $12.9$
          &
          & $10.82$
          &
          & $4.91$
          & $8.65$
          &
          & 16.71      &        \\
$\Lambda_b \rightarrow \Lambda_c K^{\ast-}$             &$0.37^{+0.18+0.12}_{-0.16-0.10}$
          & $ 0.63_{-0.26}^{+0.37} $
         &&&&&
          & $0.27$
          & $0.441$
          &
          & 0.87      &         \\
$\Lambda_b \rightarrow \Lambda_c D^{\ast-}$              &$0.82^{+0.25+0.27}_{-0.24-0.22}$
          & $ 0.84_{-0.36}^{+0.51} $ %& $ 0.89_{-0.37}^{+0.52} $
          &&&&&
          & $0.49$
          & $0.520$
          &
          & 1.38      &         \\
\hline\hline
\end{tabular}
\end{center}
\end{table*}

\begin{table*}[htbp]
\begin{center}
\caption{\label{result} Predicted decay properties of the $\Lambda_b
\rightarrow \Lambda_c^{(*)} h^-$ processes. The first error of our predictions comes from the baryon HO parameter $\alpha_{\rho}$
with uncertainties of $\pm 10\%$. The second error of our predictions comes from the meson HO parameter $\beta$
with uncertainties of $\pm 10\%$. $\mathcal{B}$ stands for the branching fraction, and $M_f$ stands for the mass of the final state
$\Lambda_c^{(*)}$. For the branching fractions ratios, the uncertainties come from $\beta$ are tiny, which are omitted.  }
\begin{tabular}{lccccccccccccccccccccccccccccccccccccccccccccc}\hline\hline
State                                                    ~~~~& Mass        ~~~~&$\mathcal{B}_{\pi^-}(10^{-3})$   ~~~~&$\frac{\mathcal{B}[\Lambda_b^0 \to \Lambda_c^{(\ast)+} \pi^-]}{\mathcal{B}[\Lambda_b^0\rightarrow \Lambda_c^+\pi^-]}$     ~~~~~&$\mathcal{B}(10^{-3})$~\cite{Chua:2019yqh}     ~~~~&$\mathcal{B}_{\rho^-}(10^{-3})$   ~~~~&$\frac{\mathcal{B}[\Lambda_b^0 \to \Lambda_c^{(\ast)+} \rho^-]}{\mathcal{B}[\Lambda_b^0\rightarrow \Lambda_c^+\rho^-]}$    ~~~~~~~&$\mathcal{B}(10^{-3})$~\cite{Chua:2019yqh}     \\\hline
%$|\Lambda_c ~~1^2S \frac{1}{2}^+ \rangle$               ~~~~& 2286                          ~~~~&$5.67^{+3.17+1.88}_{-2.56-1.53}$          ~~~~&$1.00^{+0.00}_{-0.00}$         ~~~~~&$4.16^{+2.43}_{-1.73}$    ~~~~&$6.91^{+3.08+1.62}_{-3.39-2.24}$        ~~~~&$1.00^{+0.00}_{-0.00}$       ~~~~&$12.28^{+7.19}_{-5.11}$                 \\
$|\Lambda_c ~~1^2P_{\lambda} \frac{1}{2}^- \rangle$      ~~~~& 2592                          ~~~~&$5.12^{+1.36+1.70}_{-1.52-1.39}$          ~~~~&$0.90^{-0.17}_{+0.25}$         ~~~~~&$1.09^{+0.76}_{-0.51}$    ~~~~&$5.74^{+1.32+1.90}_{-1.54-1.56}$        ~~~~&$0.90^{-0.19}_{+0.30}$       ~~~~&$2.99^{+2.20}_{-1.44}$                  \\
$|\Lambda_c ~~1^2P_{\lambda} \frac{3}{2}^- \rangle$      ~~~~& 2628                          ~~~~&$8.73^{+1.92+2.88}_{-2.37-2.37}$          ~~~~&$1.54^{-0.33}_{+0.51}$         ~~~~~&$2.40^{+4.09}_{-1.82}$    ~~~~&$8.88^{+1.71+2.94}_{-2.18-2.41}$        ~~~~&$1.39^{-0.33}_{+0.52}$       ~~~~&$4.38^{+6.78}_{-3.17}$                   \\
$|\Lambda_c ~~2^2S_{\lambda} \frac{1}{2}^+ \rangle$      ~~~~& 2767                          ~~~~&$5.73^{+0.28+1.90}_{-0.70-1.55}$          ~~~~&$1.01^{-0.33}_{+0.60}$         ~~~~~&$1.70^{+0.69}_{-0.52}$    ~~~~&$5.30^{+0.12+1.75}_{-0.49-1.44}$        ~~~~&$0.83^{-0.28}_{+0.54}$       ~~~~&$4.84^{+2.01}_{-1.50}$                  \\
$|\Lambda_c ~~1^2D_{\lambda} \frac{3}{2}^+ \rangle$      ~~~~& 2856                          ~~~~&$4.09^{+0.10+1.36}_{-0.40-1.11}$          ~~~~&$0.72^{-0.25}_{+0.47}$         ~~~~~&                          ~~~~&$4.15^{-0.01+1.38}_{-0.25-1.12}$        ~~~~&$0.65^{-0.23}_{+0.46}$       ~~~~&                                   \\
$|\Lambda_c ~~1^2D_{\lambda} \frac{5}{2}^+ \rangle$      ~~~~& 2882                          ~~~~&$4.56^{-0.16+1.51}_{-0.21-1.24}$          ~~~~&$0.80^{-0.31}_{+0.59}$         ~~~~~&                          ~~~~&$4.82^{-0.26+1.59}_{-0.07-1.31}$        ~~~~&$0.75^{-0.30}_{+0.60}$       ~~~~&                                   \\
$|\Lambda_c ~~2^2P_{\lambda} \frac{1}{2}^- \rangle$      ~~~~& 2983~\cite{Ebert:2011kk}      ~~~~&$2.39^{-0.07+1.06}_{+0.35-0.50}$          ~~~~&$0.42^{-0.18}_{+0.40}$         ~~~~~&$0.68^{+0.21}_{-0.21}$    ~~~~&$2.28^{-0.30+0.75}_{+0.24-0.62}$        ~~~~&$0.35^{-0.16}_{+0.36}$       ~~~~&$1.85^{+0.63}_{-0.60}$              \\
$|\Lambda_c ~~2^2P_{\lambda} \frac{3}{2}^- \rangle$      ~~~~& 3005~\cite{Ebert:2011kk}      ~~~~&$3.91^{-0.26+1.72}_{+0.73-0.83}$          ~~~~&$0.69^{-0.31}_{+0.71}$         ~~~~~&$1.00^{+2.00}_{-0.83}$    ~~~~&$3.79^{-0.60+1.26}_{+0.55-1.02}$        ~~~~&$0.59^{-0.27}_{+0.64}$       ~~~~&$1.93^{+3.19}_{-1.43}$                \\
$|\Lambda_c ~~1^2F_{\lambda} \frac{5}{2}^- \rangle$      ~~~~& 3097~\cite{Ebert:2011kk}      ~~~~&$1.06^{-0.21+0.35}_{+0.21-0.29}$          ~~~~&$0.19^{-0.09}_{+0.22}$         ~~~~~&                          ~~~~&$0.96^{-0.21+0.32}_{+0.23-0.26}$        ~~~~&$0.15^{-0.075}_{+0.19}$      ~~~~&                                     \\
$|\Lambda_c ~~1^2F_{\lambda} \frac{7}{2}^- \rangle$      ~~~~& 3078~\cite{Ebert:2011kk}      ~~~~&$1.46^{-0.29+0.48}_{+0.28-0.40}$          ~~~~&$0.26^{-0.12}_{+0.30}$         ~~~~~&                          ~~~~&$1.33^{-0.29+0.45}_{+0.32-0.36}$        ~~~~&$0.21^{-0.10}_{+0.26}$       ~~~~&                                       \\
$|\Lambda_c ~~3^2S_{\lambda} \frac{1}{2}^+ \rangle$      ~~~~& 3130~\cite{Ebert:2011kk}      ~~~~&$1.11^{-0.27+0.37}_{+0.33-0.30}$          ~~~~&$0.20^{-0.10}_{+0.27}$         ~~~~~&                          ~~~~&$0.83^{-0.21+0.28}_{+0.28-0.22}$        ~~~~&$0.13^{-0.068}_{+0.19}$      ~~~~&                                       \\
$|\Lambda_c ~~2^2D_{\lambda} \frac{3}{2}^+ \rangle$      ~~~~& 3189~\cite{Ebert:2011kk}      ~~~~&$1.07^{-0.27+0.36}_{+0.35-0.29}$          ~~~~&$0.19^{-0.10}_{+0.27}$         ~~~~~&                          ~~~~&$0.89^{-0.23+0.30}_{+0.32-0.24}$        ~~~~&$0.14^{-0.074}_{+0.21}$      ~~~~&                                     \\
$|\Lambda_c ~~2^2D_{\lambda} \frac{5}{2}^+ \rangle$      ~~~~& 3209~\cite{Ebert:2011kk}      ~~~~&$1.13^{-0.33+0.38}_{+0.45-0.31}$          ~~~~&$0.20^{-0.11}_{+0.31}$         ~~~~~&                          ~~~~&$0.97^{-0.29+0.33}_{+0.42-0.26}$        ~~~~&$0.15^{-0.084}_{+0.24}$      ~~~~&                                      \\
\hline\hline
State                                                    ~~~~& Mass          ~~~~&$\mathcal{B}_{K^-}(10^{-4})$   &$\frac{\mathcal{B}[\Lambda_b^0 \to \Lambda_c^{(\ast)+} K^-]}{\mathcal{B}[\Lambda_b^0\rightarrow \Lambda_c^+K^-]}$   ~~~~~&$\mathcal{B}(10^{-4})$~\cite{Chua:2019yqh}       ~~~~&$\mathcal{B}_{K^{\ast-}}(10^{-4})$   ~~~~&$\frac{\mathcal{B}[\Lambda_b^0 \to \Lambda_c^{(\ast)+} K^{\ast-}]}{\mathcal{B}[\Lambda_b^0\rightarrow \Lambda_c^+K^{\ast-}]}$  ~~~~~~~&$\mathcal{B}(10^{-4})$~\cite{Chua:2019yqh}   \\\hline
%$|\Lambda_c ~~1^2S \frac{1}{2}^+ \rangle$               ~~~~& 2286                          ~~~~&$3.11^{+1.69+1.03}_{-1.38-0.84}$         ~~~~&$1.00^{+0.00}_{-0.00}$        ~~~~~&$3.1^{+1.8}_{-1.3}$       ~~~~&$3.66^{+1.84+1.21}_{-1.55-0.99}$         ~~~~&$1.00^{+0.00}_{-0.00}$         ~~~~&$6.3^{+3.7}_{-2.6}$                               \\
$|\Lambda_c ~~1^2P_{\lambda} \frac{1}{2}^- \rangle$      ~~~~& 2592                          ~~~~&$2.77^{+0.70+0.91}_{-0.80-0.75}$         ~~~~&$0.89^{-0.17}_{+0.25}$        ~~~~~&$0.8^{+0.6}_{-0.4}$       ~~~~&$2.97^{+0.66+0.99}_{-0.77-0.80}$         ~~~~&$0.81^{-0.15}_{+0.23}$         ~~~~&$1.5^{+1.1}_{-0.7}$                         \\
$|\Lambda_c ~~1^2P_{\lambda} \frac{3}{2}^- \rangle$      ~~~~& 2628                          ~~~~&$4.65^{+0.96+1.53}_{-1.21-1.26}$         ~~~~&$1.49^{-0.33}_{+0.49}$        ~~~~~&$1.7^{+3.0}_{-1.3}$       ~~~~&$4.72^{+0.86+1.57}_{-1.11-1.28}$         ~~~~&$1.29^{-0.28}_{+0.42}$         ~~~~&$2.2^{+3.3}_{-1.6}$                              \\
$|\Lambda_c ~~2^2S_{\lambda} \frac{1}{2}^+ \rangle$      ~~~~& 2767                          ~~~~&$3.00^{+0.11+0.99}_{-0.34-0.82}$         ~~~~&$0.96^{-0.31}_{+0.57}$        ~~~~~&$1.3^{+0.05}_{-0.04}$     ~~~~&$2.76^{+0.05+0.92}_{-0.22-0.75}$         ~~~~&$0.76^{-0.25}_{+0.45}$         ~~~~&$2.5^{+0.10}_{-0.08}$                  \\
$|\Lambda_c ~~1^2D_{\lambda} \frac{3}{2}^+ \rangle$      ~~~~& 2856                          ~~~~&$2.15^{+0.03+0.71}_{-0.19-0.58}$         ~~~~&$0.69^{-0.24}_{+0.44}$        ~~~~~&                          ~~~~&$2.09^{-0.02+0.69}_{-0.11-0.57}$         ~~~~&$0.57^{-0.20}_{+0.37}$         ~~~~ &                               \\
$|\Lambda_c ~~1^2D_{\lambda} \frac{5}{2}^+ \rangle$      ~~~~& 2882                          ~~~~&$2.33^{-0.10+0.77}_{-0.07-0.63}$         ~~~~&$0.75^{-0.29}_{+0.56}$        ~~~~~&                          ~~~~&$2.41^{-0.15+0.80}_{-0.01-0.65}$         ~~~~&$0.66^{-0.25}_{+0.48}$         ~~~~&                                \\
$|\Lambda_c ~~2^2P_{\lambda} \frac{1}{2}^- \rangle$      ~~~~& 2983~\cite{Ebert:2011kk}      ~~~~&$1.22^{-0.14+0.40}_{+0.10-0.33}$         ~~~~&$0.39^{-0.17}_{+0.37}$        ~~~~~&$0.5^{+0.2}_{-0.2}$       ~~~~&$1.12^{-0.15+0.37}_{+0.13-0.31}$         ~~~~&$0.31^{-0.13}_{+0.29}$         ~~~~&$0.9^{+0.3}_{-0.3}$                              \\
$|\Lambda_c ~~2^2P_{\lambda} \frac{3}{2}^- \rangle$      ~~~~& 3005~\cite{Ebert:2011kk}      ~~~~&$1.96^{-0.29+0.65}_{+0.25-0.53}$         ~~~~&$0.63^{-0.28}_{+0.64}$        ~~~~~&$0.7^{+1.4}_{-0.6}$       ~~~~&$1.85^{-0.30+0.61}_{+0.29-0.50}$         ~~~~&$0.51^{-0.22}_{+0.51}$         ~~~~&$1.0^{1.5}_{-0.7}$                                  \\
$|\Lambda_c ~~1^2F_{\lambda} \frac{5}{2}^- \rangle$      ~~~~& 3097~\cite{Ebert:2011kk}      ~~~~&$0.53^{-0.11+0.17}_{+0.11-0.15}$         ~~~~&$0.17^{-0.08}_{+0.20}$        ~~~~~&                          ~~~~&$0.46^{-0.10+0.16}_{+0.12-0.12}$         ~~~~&$0.13^{-0.061}_{+0.15}$        ~~~~&                                  \\
$|\Lambda_c ~~1^2F_{\lambda} \frac{7}{2}^- \rangle$      ~~~~& 3078~\cite{Ebert:2011kk}      ~~~~&$0.72^{-0.15+0.24}_{+0.15-0.19}$         ~~~~&$0.23^{-0.11}_{+0.27}$        ~~~~~&                          ~~~~&$0.64^{-0.14+0.21}_{+0.16-0.17}$         ~~~~&$0.18^{-0.085}_{+0.20}$        ~~~~&                                   \\
$|\Lambda_c ~~3^2S_{\lambda} \frac{1}{2}^+ \rangle$      ~~~~& 3130~\cite{Ebert:2011kk}      ~~~~&$0.54^{-0.13+0.18}_{+0.17-0.15}$         ~~~~&$0.17^{-0.09}_{+0.23}$        ~~~~~&                          ~~~~&$0.41^{-0.11+0.13}_{+0.14-0.11}$         ~~~~&$0.11^{-0.056}_{+0.15}$        ~~~~&                                       \\
$|\Lambda_c ~~2^2D_{\lambda} \frac{3}{2}^+ \rangle$      ~~~~& 3189~\cite{Ebert:2011kk}      ~~~~&$0.53^{-0.14+0.17}_{+0.17-0.15}$         ~~~~&$0.17^{-0.09}_{+0.23}$        ~~~~~&                          ~~~~&$0.42^{-0.11+0.14}_{+0.15-0.11}$         ~~~~&$0.12^{-0.059}_{+0.16}$        ~~~~&                                  \\
$|\Lambda_c ~~2^2D_{\lambda} \frac{5}{2}^+ \rangle$      ~~~~& 3209~\cite{Ebert:2011kk}      ~~~~&$0.54^{-0.17+0.17}_{+0.22-0.15}$         ~~~~&$0.17^{-0.09}_{+0.27}$        ~~~~~&                          ~~~~&$0.45^{-0.14+0.15}_{+0.20-0.12}$         ~~~~&$0.12^{-0.067}_{+0.19}$        ~~~~&                                  \\
\hline\hline
State                                                    ~~~~& Mass         ~~~~&$\mathcal{B}_{D^-}(10^{-4})$   ~~~~&$\frac{\mathcal{B}[\Lambda_b^0 \to \Lambda_c^{(\ast)+} D^-]}{\mathcal{B}[\Lambda_b^0\rightarrow \Lambda_c^+D^-]}$ ~~~~~&$\mathcal{B}$~\cite{Chua:2019yqh}      ~~~~&$\mathcal{B}_{D^{\ast-}}(10^{-4})$   ~~~~&$\frac{\mathcal{B}[\Lambda_b^0 \to \Lambda_c^{(\ast)+} D^{\ast-}]}{\mathcal{B}[\Lambda_b^0\rightarrow \Lambda_c^+D^{\ast-}]}$   ~~~~~~~&$\mathcal{B}(10^{-4})$~\cite{Chua:2019yqh}  \\\hline
%$|\Lambda_c ~~1^2S \frac{1}{2}^+ \rangle$               ~~~~& 2286                          ~~~~&$4.30^{+1.43+1.43}_{-1.38-1.16}$        ~~~~&$1.00^{+0.00}_{-0.00}$        ~~~~~&$4.7^{+3.0}_{-0.21}$       ~~~~&$8.15^{+2.47+2.70}_{-2.45-2.21}$        ~~~~&$1.00^{+0.00}_{-0.00}$           ~~~~&$8.4^{+5.1}_{3.6}$             \\
$|\Lambda_c ~~1^2P_{\lambda} \frac{1}{2}^- \rangle$      ~~~~& 2592                          ~~~~&$2.47^{+0.26+0.82}_{-0.36-0.67}$        ~~~~&$0.57^{-0.098}_{+0.15}$       ~~~~~&$0.7^{+0.7}_{-0.4}$        ~~~~&$4.07^{+0.32+1.35}_{-0.47-1.10}$        ~~~~&$0.50^{-0.087}_{+0.13}$          ~~~~&$1.2^{+1.1}_{-0.7}$                   \\
$|\Lambda_c ~~1^2P_{\lambda} \frac{3}{2}^- \rangle$      ~~~~& 2625                          ~~~~&$3.59^{+0.13+1.19}_{-0.32-0.97}$        ~~~~&$0.83^{-0.19}_{+0.29}$        ~~~~~&$1.3^{+2.2}_{1.0}$         ~~~~&$3.42^{+0.083+1.13}_{-0.24-0.93}$       ~~~~&$0.42^{-0.090}_{+0.14}$          ~~~~&$1.3^{+1.7}_{-0.8}$                     \\
$|\Lambda_c ~~2^2S_{\lambda} \frac{1}{2}^+ \rangle$      ~~~~& 2767                          ~~~~&$1.49^{-0.11+0.49}_{+0.081-0.40}$       ~~~~&$0.35^{-0.11}_{+0.19}$        ~~~~~&$1.5^{+0.7}_{-0.5}$        ~~~~&$1.21^{-0.108+0.40}_{+0.097-0.33}$      ~~~~&$0.15^{-0.045}_{+0.081}$         ~~~~&$2.6^{+1.2}_{-0.9}$                  \\
$|\Lambda_c ~~1^2D_{\lambda} \frac{3}{2}^+ \rangle$      ~~~~& 2856                          ~~~~&$0.94^{-0.087+0.31}_{+0.078-0.26}$      ~~~~&$0.22^{-0.070}_{+0.13}$       ~~~~~&                           ~~~~&$1.25^{-0.141+0.42}_{+0.15-0.34}$       ~~~~&$0.15^{-0.049}_{+0.092}$         ~~~~&                                     \\
$|\Lambda_c ~~1^2D_{\lambda} \frac{5}{2}^+ \rangle$      ~~~~& 2882                          ~~~~&$0.77^{-0.14+0.25}_{+0.15-0.21}$        ~~~~&$0.18^{-0.069}_{+0.14}$       ~~~~~&                           ~~~~&$1.03^{-0.20+0.34}_{+0.24-0.28}$        ~~~~&$0.13^{-0.049}_{+0.096}$         ~~~~&                                   \\
$|\Lambda_c ~~2^2P_{\lambda} \frac{1}{2}^- \rangle$      ~~~~& 2983~\cite{Ebert:2011kk}      ~~~~&$0.30^{-0.053+0.10}_{+0.069-0.082}$     ~~~~&$0.07^{-0.027}_{+0.057}$      ~~~~~&$0.4^{+0.2}_{-0.2}$        ~~~~&$0.35^{-0.064+0.11}_{+0.087-0.094}$     ~~~~&$0.042^{-0.016}_{+0.034}$        ~~~~&$0.6^{+0.3}_{-0.3}$                    \\
$|\Lambda_c ~~2^2P_{\lambda} \frac{3}{2}^- \rangle$      ~~~~& 3005~\cite{Ebert:2011kk}      ~~~~&$0.39^{-0.090+0.13}_{+0.12-0.11}$       ~~~~&$0.09^{-0.038}_{+0.085}$      ~~~~~&$0.7^{+1.0}_{-0.5}$        ~~~~&$0.44^{-0.10+0.15}_{+0.15-0.12}$        ~~~~&$0.054^{-0.022}_{+0.049}$        ~~~~&$0.6^{+0.6}_{-0.3}$                     \\
$|\Lambda_c ~~1^2F_{\lambda} \frac{5}{2}^- \rangle$      ~~~~& 3097~\cite{Ebert:2011kk}      ~~~~&$0.062^{-0.018+0.020}_{+0.028-0.017}$   ~~~~&$0.014^{-0.007}_{+0.016}$     ~~~~~&                           ~~~~&$0.056^{-0.017+0.019}_{+0.028-0.015}$   ~~~~&$0.0069^{-0.0033}_{+0.0079}$     ~~~~&                           \\
$|\Lambda_c ~~1^2F_{\lambda} \frac{7}{2}^- \rangle$      ~~~~& 3078~\cite{Ebert:2011kk}      ~~~~&$0.090^{-0.027+0.030}_{+0.040-0.024}$   ~~~~&$0.021^{-0.010}_{+0.024}$     ~~~~~&                           ~~~~&$0.084^{-0.026+0.028}_{+0.041-0.023}$   ~~~~&$0.010^{-0.0049}_{+0.012}$       ~~~~&                         \\
$|\Lambda_c ~~3^2S_{\lambda} \frac{1}{2}^+ \rangle$      ~~~~& 3130~\cite{Ebert:2011kk}      ~~~~&$0.056^{-0.015+0.019}_{+0.024-0.015}$   ~~~~&$0.013^{-0.006}_{+0.014}$     ~~~~~&                           ~~~~&$0.029^{-0.008+0.010}_{+0.012-0.008}$   ~~~~&$0.0035^{-0.0015}_{+0.0037}$     ~~~~&                                       \\
$|\Lambda_c ~~2^2D_{\lambda} \frac{3}{2}^+ \rangle$      ~~~~& 3189~\cite{Ebert:2011kk}      ~~~~&$0.043^{-0.012+0.014}_{+0.019-0.012}$   ~~~~&$0.010^{-0.005}_{+0.011}$     ~~~~~&                           ~~~~&$0.033^{-0.009+0.011}_{+0.014-0.009}$   ~~~~&$0.0040^{-0.0018}_{+0.0043}$     ~~~~&                      \\
$|\Lambda_c ~~2^2D_{\lambda} \frac{5}{2}^+ \rangle$      ~~~~& 3209~\cite{Ebert:2011kk}      ~~~~&$0.028^{-0.010+0.009}_{+0.017-0.008}$   ~~~~&$0.006^{-0.003}_{+0.009}$     ~~~~~&                           ~~~~&$0.020^{-0.007+0.007}_{+0.012-0.005}$   ~~~~&$0.0025^{-0.0012}_{+0.0032}$     ~~~~&                       \\
\hline\hline
State                                                    ~~~~& Mass         ~~~~&$\mathcal{B}_{D_s^-}(10^{-3})$   ~~~~&$\frac{\mathcal{B}[\Lambda_b^0 \to \Lambda_c^{(\ast)+} D_s^-]}{\mathcal{B}[\Lambda_b^0\rightarrow \Lambda_c^+D_s^-]}$ ~~~~~&$\mathcal{B}(10^{-3})$~\cite{Chua:2019yqh}     ~~~~&$\mathcal{B}_{D_s^{\ast-}}(10^{-3})$   ~~~~&$\frac{\mathcal{B}[\Lambda_b^0 \to \Lambda_c^{(\ast)+} D_s^{\ast-}]}{\mathcal{B}[\Lambda_b^0\rightarrow \Lambda_c^+D_s^{\ast-}]}$  ~~~~~~~&$\mathcal{B}(10^{-3})$~\cite{Chua:2019yqh}   \\\hline
%$|\Lambda_c ~~1^2S \frac{1}{2}^+ \rangle$               ~~~~& 2286                          ~~~~&$9.75^{+3.04+3.23}_{-3.00-2.64}$         ~~~~&$1.00^{+0.00}_{-0.00}$           ~~~~~&$11.92^{+7.69}_{-5.28}$    ~~~~&$17.59^{+4.93+5.82}_{-4.99-4.77}$          ~~~~&$1.00^{+0.00}_{-0.00}$               ~~~~&$17.49^{10.60}_{-7.48}$                       \\
$|\Lambda_c ~~1^2P_{\lambda} \frac{1}{2}^- \rangle$      ~~~~& 2592                          ~~~~&$5.37^{+0.50+1.78}_{-0.70-1.45}$         ~~~~&$0.55^{-0.092}_{+0.14}$          ~~~~~&$1.72^{+1.71}_{-1.01}$     ~~~~&$8.23^{+0.54+2.73}_{-0.82-2.23}$           ~~~~&$0.47^{-0.078}_{+0.12}$              ~~~~&$2.28^{+2.21}_{-1.29}$                        \\
$|\Lambda_c ~~1^2P_{\lambda} \frac{3}{2}^- \rangle$      ~~~~& 2625                          ~~~~&$7.52^{+0.14+2.49}_{-0.52-2.04}$         ~~~~&$0.77^{-0.17}_{+0.27}$           ~~~~~&$2.88^{+4.92}_{-2.16}$     ~~~~&$7.02^{+0.046+2.32}_{-0.34-1.90}$          ~~~~&$0.40^{-0.085}_{+0.13}$              ~~~~&$2.41^{+2.98}_{-1.52}$                      \\
$|\Lambda_c ~~2^2S_{\lambda} \frac{1}{2}^+ \rangle$      ~~~~& 2767                          ~~~~&$2.95^{-0.25+0.98}_{+0.21-0.80}$         ~~~~&$0.30^{-0.091}_{+0.17}$          ~~~~~&$3.54^{+1.73}_{-1.24}$     ~~~~&$2.29^{-0.22+0.76}_{+0.22-0.62}$           ~~~~&$0.13^{-0.038}_{+0.069}$             ~~~~&$5.29^{+2.54}_{-1.84}$                  \\
$|\Lambda_c ~~1^2D_{\lambda} \frac{3}{2}^+ \rangle$      ~~~~& 2856                          ~~~~&$1.85^{-0.18+0.61}_{+0.18-0.50}$         ~~~~&$0.19^{-0.059}_{+0.11}$          ~~~~~&                           ~~~~&$2.19^{-0.25+0.73}_{+0.28-0.59}$           ~~~~&$0.12^{-0.039}_{+0.072}$             ~~~~&                                           \\
$|\Lambda_c ~~1^2D_{\lambda} \frac{5}{2}^+ \rangle$      ~~~~& 2882                          ~~~~&$1.40^{-0.28+0.46}_{+0.32-0.38}$         ~~~~&$0.14^{-0.056}_{+0.11}$          ~~~~~&                           ~~~~&$1.65^{-0.35+0.55}_{+0.43-0.45}$           ~~~~&$0.094^{-0.036}_{+0.072}$            ~~~~&                                           \\
$|\Lambda_c ~~2^2P_{\lambda} \frac{1}{2}^- \rangle$      ~~~~& 2983~\cite{Ebert:2011kk}      ~~~~&$0.53^{-0.09+0.18}_{+0.13-0.14}$         ~~~~&$0.055^{-0.020}_{+0.043}$        ~~~~~&$0.87^{+0.46}_{-0.38}$     ~~~~&$0.53^{-0.095+0.18}_{+0.13-0.14}$          ~~~~&$0.030^{-0.011}_{+0.023}$            ~~~~&$1.16^{+0.62}_{-0.48}$                        \\
$|\Lambda_c ~~2^2P_{\lambda} \frac{3}{2}^- \rangle$      ~~~~& 3005~\cite{Ebert:2011kk}      ~~~~&$0.65^{-0.15+0.22}_{+0.22-0.18}$         ~~~~&$0.067^{-0.028}_{+0.062}$        ~~~~~&$1.69^{+2.30}_{-1.23}$     ~~~~&$0.63^{-0.15+0.21}_{+0.22-0.17}$           ~~~~&$0.036^{-0.014}_{+0.031}$            ~~~~&$1.11^{+1.07}_{-0.62}$                         \\
$|\Lambda_c ~~1^2F_{\lambda} \frac{5}{2}^- \rangle$      ~~~~& 3097~\cite{Ebert:2011kk}      ~~~~&$0.093^{-0.028+0.031}_{+0.044-0.025}$    ~~~~&$0.010^{-0.004}_{+0.011}$        ~~~~~&                           ~~~~&$0.066^{-0.021+0.022}_{+0.034-0.018}$      ~~~~&$0.0037^{-0.0017}_{+0.0042}$         ~~~~&                                           \\
$|\Lambda_c ~~1^2F_{\lambda} \frac{7}{2}^- \rangle$      ~~~~& 3078~\cite{Ebert:2011kk}      ~~~~&$0.14^{-0.042+0.046}_{+0.065-0.037}$     ~~~~&$0.014^{-0.007}_{+0.016}$        ~~~~~&                           ~~~~&$0.10^{-0.032+0.034}_{+0.051-0.027}$       ~~~~&$0.0058^{-0.0027}_{+0.0064}$         ~~~~&                                           \\
$|\Lambda_c ~~3^2S_{\lambda} \frac{1}{2}^+ \rangle$      ~~~~& 3130~\cite{Ebert:2011kk}      ~~~~&$0.083^{-0.022+0.027}_{+0.035-0.022}$    ~~~~&$0.008^{-0.004}_{+0.009}$        ~~~~~&                           ~~~~&$0.036^{-0.0091+0.012}_{+0.015-0.0099}$    ~~~~&$0.0021^{-0.0009}_{+0.0020}$         ~~~~&                                       \\
$|\Lambda_c ~~2^2D_{\lambda} \frac{3}{2}^+ \rangle$      ~~~~& 3189~\cite{Ebert:2011kk}      ~~~~&$0.061^{-0.016+0.020}_{+0.026-0.017}$    ~~~~&$0.006^{-0.003}_{+0.007}$        ~~~~~&                           ~~~~&$0.035^{-0.0087+0.012}_{+0.014-0.0096}$    ~~~~&$0.0020^{-0.0008}_{+0.0019}$         ~~~~&                                             \\
$|\Lambda_c ~~2^2D_{\lambda} \frac{5}{2}^+ \rangle$      ~~~~& 3209~\cite{Ebert:2011kk}      ~~~~&$0.035^{-0.012+0.011}_{+0.022-0.009}$    ~~~~&$0.004^{-0.002}_{+0.005}$        ~~~~~&                           ~~~~&$0.018^{-0.0062+0.0059}_{+0.011-0.0049}$   ~~~~&$0.0010^{-0.0005}_{+0.0013}$         ~~~~&                                          \\
\hline\hline
\end{tabular}
\end{center}
\end{table*}

\subsection{$\Lambda_b \rightarrow \Lambda_c h^-$}

Firstly, we study the weak decay process $\Lambda_b \rightarrow \Lambda_c h^-$ as a test of our model.
Our predicted branching fractions for each decay channel are given in Table~\ref{result1S}. For a comparison, the experimental data and some results
from other theoretical models are listed in the same table as well.
From Table~\ref{result1S}, it is seen that our predicted central values of branching fractions for $\Lambda_b  \rightarrow \Lambda_c (\pi^-, K^-, D^-, D_s^-, D_s^{*-})$,  $(5.67, 0.31, 0.43, 10.3, 17.5)\times 10^{-3}$, are in good agreement with the experimental data~\cite{pdg}. Furthermore, the predicted branching fraction ratios
\begin{eqnarray}
R_{K/\pi} &=& \frac{\mathcal{B}[\Lambda_b \to \Lambda_c K^{-}]}{\mathcal{B}[\Lambda_b \to \Lambda_c  \pi^{-}]}\simeq 0.0548^{-0.0006}_{+0.0007} \pm 0.0,  \\
R_{D/D_s} &=& \frac{\mathcal{B}[\Lambda_b \to \Lambda_c D^{-}]}{\mathcal{B}[\Lambda_b \to \Lambda_c  D_s^{-}]}\simeq0.0440^{+0.0007-0.011}_{-0.0009+0.017}, \\
R_{D^*_s/D_s} &=& \frac{\mathcal{B}[\Lambda_b \to \Lambda_c D_s^{*-}]}{\mathcal{B}[\Lambda_b \to \Lambda_c  D_s^{-}]}\simeq1.80^{-0.04}_{+0.06} \pm 0.0,
\end{eqnarray}
are also consistent with the measured central values, $R_{K/\pi}^{exp}=0.0731\pm 0.0022$, $R_{D/D_s}^{exp}=0.042\pm 0.006$, $R_{D^*_s/D_s}^{exp}=1.668\pm 0.022^{+0.061}_{-0.055}$,, respectively.

For the processes $\Lambda_b  \rightarrow \Lambda_c (\rho^-, K^{\ast-}, D^{\ast-})$, which have not been measured in experiments, the branching
fractions are predicted to be $(6.91^{+3.08+1.62}_{-3.39-2.24}, 0.37^{+0.18+0.12}_{-0.16-0.10}, 0.82^{+0.25+0.27}_{-0.24-0.22})$ $\times 10^{-3}$. From Table~\ref{result1S}, it is found that there are obviously model dependencies in the various model predictions. For example, for the $\Lambda_b  \rightarrow \Lambda_c \rho^-$ process our predicted branching
fractions are roughly consistent with those of heavy quark
effective theory~\cite{Mannel:1992ti}, covariant oscillator quark model~\cite{Mohanta:1998iu}, covariant light-front quark model~\cite{Zhu:2018jet}, however, about a factor of $\sim 2$ smaller than those predicted in Refs.~\cite{Chua:2019yqh,Ke:2019smy,Giri:1997te}.

To further understand the branching fractions of the vector meson emission processes, $\mathcal{B}(\Lambda_b  \rightarrow \Lambda_c V)$ $(V=\rho^-, K^{\ast-},D^*,D_s^*)$, we analyze their transition amplitudes. The extracted transition amplitudes for the $\Lambda_b  \rightarrow \Lambda_c V$ process are given by
\begin{eqnarray}\label{equa1}
\mathcal{M}^{1/2,1/2,1,PV}_{-1/2,-1/2,0}&&=  32\sqrt{3} \pi ^{3/4}\mathcal{C}_W  \beta^{9/2} T_\alpha F(q^2),\\
\mathcal{M}^{1/2,1/2,1,PV}_{-1/2,1/2,-1}&&= - 16 \sqrt{6} \pi^{3/4}\mathcal{C}_W   \beta^{9/2}  T_\alpha F(q^2)\label{equa2},
\end{eqnarray}
\begin{eqnarray}\label{equa3}
\mathcal{M}^{1/2,1/2,1,PC}_{-1/2,-1/2,0}&&= -\mathcal{C}_W
T_\alpha |\textbf{q}| F(q^2) \Bigg[\alpha_{\lambda_i}^2 \left(2-\frac{4m_2}{m_3}+\frac{2 m_2}{m_4}+\frac{2 m_2}{m_5}+\frac{m_{3'}}{m_4}+\frac{m_{3'}}{m_5}\right)\nonumber  \\
&&+\alpha_{\lambda_f}^2 \left(2+\frac{2 m_2}{m_4}+\frac{2 m_2}{m_5}+\frac{4 m_2}{m_{3'}}+\frac{m_{3'}}{m_4}+\frac{m_{3'}}{m_5}\right)\Bigg],
\end{eqnarray}
\begin{eqnarray}\label{equa4}
\mathcal{M}^{1/2,1/2,1,PC}_{-1/2,1/2,-1}&&= -\sqrt{2}\mathcal{C}_W
T_\alpha |\textbf{q}| F(q^2) \Bigg[\alpha_{\lambda_i}^2 \left(2+\frac{4 m_2}{m_3}-\frac{2 m_2}{m_4}+\frac{2 m_2}{m_5}-\frac{m_{3'}}{m_4}
+\frac{m_{3'}}{m_5}\right) \nonumber  \\
&&+\alpha_{\lambda_f}^2 \left(2-\frac{2 m_2}{m_4}+\frac{2 m_2}{m_5}+\frac{4 m_2}{m_{3'}}-\frac{m_{3'}}{m_4}+\frac{m_{3'}}{m_5}\right)\Bigg],
\end{eqnarray}
with
\begin{eqnarray}
\mathcal{C}_W &=& \frac{G_F}{\sqrt{2}}\frac{1}{8\pi^{3}}V_{cb}V_{45}^{\ast}\left(C_1+\frac{C_2}{3}\right),\\
F(q^2)&=&\exp \left(-\frac{3 m_2^2 q^2}{\left(\alpha_{\lambda_f}^2+\alpha_{\lambda_i}^2\right) (2 m_2+m_3')^2}\right),\\
T_\alpha&=&\frac{8\sqrt{3} \pi^{3/4} \alpha_{\lambda_f}^3  \alpha_{\lambda_i}^3 \alpha_{\rho_f}^{3/2} \alpha_{\rho_i}^{3/2}\beta^{3/2}}{(2 m_2+m_{3'})\left(\alpha_{\lambda_f}^2+\alpha_{\lambda_i}^2\right)^{5/2} \left(\alpha_{\rho_f}^2+\alpha_{\rho_i}^2\right)^{3/2}},
\end{eqnarray}
where ($\alpha_{\rho_i}$, $\alpha_{\lambda_i}$) and ($\alpha_{\rho_f}$, $\alpha_{\lambda_f}$) are ($\rho$, $\lambda$)-mode HO parameters of $\Lambda_b$ and $\Lambda_c$, respectively, while $\beta$ stands for the HO parameters of the emitting mesons.
From the amplitudes given in Eqs.~(\ref{equa1})-(\ref{equa4}), one can see that the contributions of the parity-violating amplitudes $\mathcal{M}^{1/2,1/2,1,PV}_{-1/2,-1/2,0}$ and $\mathcal{M}^{1/2,1/2,1,PV}_{-1/2,1/2,-1}$ are relatively small due to the suppression of the heavy charmed quark mass, $m_{3'}$. For the $\Lambda_b  \to \Lambda_c  D^{\ast-}, \Lambda_c  D^{\ast-}_s$ processes,
the dominant contributions arise from the parity-conserving amplitudes $\mathcal{M}^{1/2,1/2,1,PC}_{-1/2,-1/2,0}$ and $\mathcal{M}^{1/2,1/2,1,PC}_{-1/2,1/2,-1}$ given in Eqs.~(\ref{equa3}) and (\ref{equa4}), where the $\frac{m_{3'}}{m_5}\simeq 3.2$ term plays
a crucial role (note that $m_{3'}=m_c$, $m_{5}=m_u$). For the $\Lambda_b  \to \Lambda_c  (\rho^-, K^{\ast-})$ processes,
the dominant contribution arise from the parity-conserving amplitude $\mathcal{M}^{1/2,1/2,1,PC}_{-1/2,-1/2,0}$, however, the $\mathcal{M}^{1/2,1/2,1,PC}_{-1/2,1/2,-1}$ contribution is strongly suppressed due to a large cancellations between the main terms $\frac{m_{2(3')}}{m_5}$ and $\frac{m_{2(3')}}{m_4}$. This cancellation mechanism is the main reason why large ratios
$\frac{\mathcal{B}(\Lambda_b  \rightarrow \Lambda_c D^{\ast-}_s)}{\mathcal{B}(\Lambda_b  \rightarrow \Lambda_c \rho^-)}\simeq 2.5^{-0.4}_{+0.6}$ and $\frac{\mathcal{B}(\Lambda_b  \rightarrow \Lambda_c D^{\ast-})}{\mathcal{B}(\Lambda_b  \rightarrow \Lambda_c K^{*-})}\simeq 2.2^{-0.3}_{+0.5}$
are predicted in this work.

Finally, it should be mentioned that for the $\Lambda_b  \rightarrow \Lambda_c (\pi^-, K^-)$ processes,
except for a sign difference, the transition amplitudes $\mathcal{M}^{1/2,1/2,0,PV}_{-1/2,-1/2,0}$ and $\mathcal{M}^{1/2,1/2,0,PC}_{-1/2,-1/2,0}$ have the same forms of those for the $\Lambda_b  \rightarrow \Lambda_c V$ process given in Eqs.~(\ref{equa1}) and (\ref{equa3}), respectively.
Since the $\mathcal{M}^{1/2,1/2,1,PV}_{-1/2,1/2,-1}$ and $\mathcal{M}^{1/2,1/2,1,PC}_{-1/2,1/2,-1}$ are negligibly small for the $\Lambda_b  \rightarrow \Lambda_c (\rho^-, K^{*-})$ processes,  we obtain $\mathcal{B}[\Lambda_b  \rightarrow \Lambda_c \pi^-] \simeq \mathcal{B}[\Lambda_b  \rightarrow \Lambda_c \rho^-]$ and $\mathcal{B}[\Lambda_b  \rightarrow \Lambda_c K^-)] \simeq \mathcal{B}[\Lambda_b  \rightarrow \Lambda_c K^{\ast-}]$ as those given in Table~\ref{result1S}.

\subsection{$\Lambda_b \rightarrow \Lambda_c(1P) h^-$ }\label{1p1b}

There are two $\lambda$-mode  $1P$-wave $\Lambda_c$ states $|\Lambda_c ~1^2P_{\lambda} \frac{1}{2}^- \rangle$ and $|\Lambda_c ~1^2P_{\lambda} \frac{3}{2}^- \rangle$ predicted within the quark model, which should correspond to the two well-established resonances $\Lambda_c(2595)^+$ and $\Lambda_c(2625)^+$ listed by PDG. In this work, we study the $\Lambda_b \rightarrow \Lambda_c(2595,2625)^+ h^-$ ($h^-=\pi^-, \rho^-, K^{-}, K^{\ast-}, D^{-}, D^{\ast-}, D_s^{-}, D_s^{\ast-}$) processes, and our results
are given in Table~\ref{result}.

It is seen that the nonleptonic weak decays associated with the emission of $\pi^-$ mesons, $\Lambda_b \rightarrow \Lambda_c(2595,2625)^+ \pi^-$, have large decay rates. The branching fractions are predicted to be
\begin{eqnarray}\label{1p}
\mathcal{B}[\Lambda_b \rightarrow \Lambda_c(2595,2625)^+ \pi^-]\simeq(5.12^{+1.36+1.70}_{-1.52-1.39},8.73^{+1.92+2.88}_{-2.37-2.37}) \times 10^{-3},
\end{eqnarray}
the order of magnitude, $\mathcal{O}(10^{-3})$, is consistent with that predicted
within the light-front quark model~\cite{Chua:2019yqh} and constituent quark model~\cite{Li:2022hcn}.
The branching fraction ratio between the $\Lambda_c(2595^+) \pi^-$ and $\Lambda_c(2625)^+ \pi^-$ is estimated to be
\begin{eqnarray}\label{1pa}
\frac{\mathcal{B}[\Lambda_b \rightarrow \Lambda_c(2595)^+ \pi^-]}{\mathcal{B}[\Lambda_b \rightarrow \Lambda_c(2625)^+ \pi^-]}\simeq0.59^{+0.02}_{-0.02},
\end{eqnarray}
which is in agreement with the prediction, $\sim0.45$, in Ref.~\cite{Chua:2019yqh}.

In experiments, the decay chains $\Lambda_b \to \Lambda_c(2595,2625)^+ \pi^-$, with $\Lambda_c(2595,2625)^+ \to \Lambda_c^+ \pi^+ \pi^-$, have been observed by the CDF and LHCb collaborations~\cite{CDF:2011aa,LHCb:2011poy}. The branching fractions relative to the $\Lambda_b\to\Lambda_c^+ \pi^-\pi^+ \pi^-$ measured by the CDF are $\mathcal{B}[\Lambda_b \to \Lambda_c(2595)^+ \pi^-]\cdot\mathcal{B}[\Lambda_c(2595)^+\to\Lambda_c^+ \pi^+ \pi^-]=(2.3\pm0.5\pm0.4)\%$ and  $\mathcal{B}[\Lambda_b \to \Lambda_c(2625)^+ \pi^-]\cdot\mathcal{B}[\Lambda_c(2625)^+\to\Lambda_c^+ \pi^+ \pi^-]=(6.8\pm1.0\pm2.3)\%$~\cite{CDF:2011aa}, which are comparable with the LHCb measured values $(4.4\pm 1.7^{+0.6}_{-0.4})\%$ and $(4.3\pm 1.5\pm 0.3)\%$, respectively. Combining the measured central value $\mathcal{B}[\Lambda_c(2625)^+\to\Lambda_c^+ \pi^+ \pi^-]=50.2\%$ and the theoretical prediction $\mathcal{B}[\Lambda_c(2595)^+\to\Lambda_c^+ \pi^+ \pi^-]=39.9\%$ in Ref.~\cite{Ponkhuha:2024gms}, from the data measured at CDF one can extract a branching fraction ratio $\frac{\mathcal{B}[\Lambda_b \rightarrow \Lambda_c(2595)^+ \pi^-]}{\mathcal{B}[\Lambda_b \rightarrow \Lambda_c(2625)^+ \pi^-]}\simeq0.42$ (central value), which is close to our prediction in Eq.~(\ref{1pa}).

Due to the large CKM matrix element $V_{cs}=0.987$, the nonleptonic weak decays associated $D_s$ and $D_s^*$ emitting, $\Lambda_b \rightarrow \Lambda_c(2595,2625)^+ D_s^-,\Lambda_c(2595,2625)^+ D_s^{*-}$, have large decay rates as well as that of the $\pi^-$ meson emitting processes.
The order of magnitude of the branching fractions can reach up to $\mathcal{O}(10^{-3})$. The $\Lambda_b \to \Lambda_c(2595,2625)^+ D_s^{(*)-}$ processes were also studied within the light-front quark model in Ref.~\cite{Chua:2019yqh}, where the predicted branching fractions are systematically a factor of $\sim 3$ smaller than ours. According to our predictions, the branching fractions of $\Lambda_b \to \Lambda_c(2595,2625)^+ D_s^{(*)-}$ are only a factor of $\sim2$ smaller those of $\Lambda_b \to \Lambda_c D_s^{(*)-}$, thus, with the accumulation of the $\Lambda_b$ sample, the $\Lambda_c(2595)^+ D_s^{(*)-}$ and $\Lambda_c(2625)^+ D_s^{(*)-}$ channels may have great potential to be observed in future experiments.

\subsection{$\Lambda_b \rightarrow \Lambda_c(1D_{\lambda}) h^-$ }\label{1p1a}

There are two $\lambda$-mode  $1D$-wave $\Lambda_c$ states $|\Lambda_c ~1^2D_{\lambda} \frac{3}{2}^+ \rangle$ and $|\Lambda_c ~1^2D_{\lambda} \frac{5}{2}^+ \rangle$ predicted within the quark model. The $\Lambda_c(2860)3/2^+$ and $\Lambda_c(2880)5/2^+$ resonances observed in experiments
favor the $\lambda$-mode $1D$-wave states~\cite{Lin:2021wrb,Kim:2020imk,Kim:2024tbf,Ponkhuha:2024gms,Arifi:2021orx,Cheng:2006dk,Cheng:2015naa,Chen:2007xf,Chen:2015kpa,
Chen:2016phw,Chen:2017sci,Ebert:2007nw,Ebert:2011kk,Chen:2009tm,Chen:2014nyo,Chen:2016iyi,Xie:2025gom,Yu:2023bxn,
Garcia-Tecocoatzi:2022zrf,Gong:2021jkb,Zhong:2007gp,Yao:2018jmc,
Chen:2017aqm,Guo:2019ytq}. With this assignment, we study the $\Lambda_b \rightarrow \Lambda_c(2860,2880)^+ h^-$ ($h^-=\pi^-, \rho^-, K^{-}, K^{\ast-}, D^{-}, D^{\ast-}, D_s^{-}, D_s^{\ast-}$) processes, and our results
are given in Table~\ref{result}. It is found that the $\Lambda_b$ has large decay rates into both the $\Lambda_c(2860)^+\pi^-$
and $\Lambda_c(2880)^+\pi^-$ final states. The branching fractions are predicted to be
\begin{eqnarray}\label{1p}
\mathcal{B}[\Lambda_b \rightarrow \Lambda_c(2860,2880)^+ \pi^-]\simeq(4.09^{+0.10+1.36}_{-0.40-1.11},4.56^{-0.16+1.51}_{-0.21-1.24}) \times 10^{-3},
\end{eqnarray}
which are comparable to that of the $\Lambda_c^+\pi^-$ channel.

The $\Lambda_b \rightarrow \Lambda_c(2860,2880)^+ \pi^-$
processes have been observed by the LHCb collaboration~\cite{LHCb:2017jym}. From the data, they extract a branching fraction
ratio
\begin{eqnarray}\label{1pd}
\frac{\mathcal{B}[\Lambda_b \to \Lambda_c(2860)^+ \pi^-]\cdot \mathcal{B}[\Lambda_c(2860)^+\to D^0p]}{\mathcal{B}[\Lambda_b \to \Lambda_c(2880)^+ \pi^-]\cdot \mathcal{B}[\Lambda_c(2880)^+\to D^0p]}\simeq 4.54^{+0.51}_{-0.39}\pm 0.12.
\end{eqnarray}
In a recent work~\cite{Zhang:2024afw}, the partial widths of $\Lambda_c(2860,2880)^+$ into the $D^0p$ channel are predicted to be $\Gamma[\Lambda_c(2860)^+\to D^0p]\simeq 22$ MeV and $\Gamma[\Lambda_c(2880)^+\to D^0p]\simeq 0.35$ MeV. Combining these predicted partial widths with the measured widths
of $\Gamma_{\Lambda_c(2860)}\simeq 67.6$ MeV and $\Gamma_{\Lambda_c(2880)}\simeq 5.6$ MeV from PDG,
we estimate the branching fractions of $\Lambda_c(2860,2880)$ into the $D^0p$ as $\mathcal{B}[\Lambda_c(2860)^+\to D^0p]\simeq 32.5\%$ and $\mathcal{B}[\Lambda_c(2880)^+\to D^0p]\simeq 6.3\%$. Further combining the measured ratio given in Eq.~(\ref{1pd}), one can extract an experimental ratio of $\frac{\mathcal{B}[\Lambda_b \to \Lambda_c(2860)^+ \pi^-]}{\mathcal{B}[\Lambda_b \to \Lambda_c(2880)^+ \pi^-]}\simeq 0.88$, which is in good agreement with the value,
\begin{eqnarray}\label{1p}
\frac{\mathcal{B}[\Lambda_b \to \Lambda_c(2860)^+ \pi^-]}{\mathcal{B}[\Lambda_b \to \Lambda_c(2880)^+ \pi^-]}\simeq 0.90^{+0.05}_{-0.05},
\end{eqnarray}
predicted in theory. Thus, the $\Lambda_c(2860)3/2^+$ and $\Lambda_c(2880)5/2^+$ production data via $\Lambda_b$ weak decays from LHCb support their $\lambda$-mode $1D$-wave assignment.

\subsection{$\Lambda_b \rightarrow \Lambda_c(2P_{\lambda}) h^-$ }\label{1p1}

There are two $\lambda$-mode $2P$-wave states, $|\Lambda_c~2^2P_{\lambda} \frac{1}{2}^- \rangle$ and $|\Lambda_c~2^2P_{\lambda} \frac{3}{2}^- \rangle$, with masses of $\sim3.0$ GeV according to various quark model predictions~\cite{Oudichhya:2023awb,Jakhad:2023mni,Kim:2021ywp,
Shah:2016nxi,Shah:2016mig,Lu:2019rtg,Yu:2022ymb,Luo:2019qkm,Azizi:2022dpn,Yang:2023fsc,Roberts:2007ni,Capstick:1986ter,Chen:2016iyi,Chen:2014nyo,Ebert:2011kk}. The weak decays of $\Lambda_b$ may provide a good opportunity to establish the $2P$-wave $\Lambda_c$ states.
In this work, we study the $\Lambda_b \rightarrow \Lambda_c(2P_{\lambda})^+ h^-$ ($h^-=\pi^-, \rho^-, K^{-}, K^{\ast-}, D^{-}, D^{\ast-}, D_s^{-}, D_s^{\ast-}$) processes, and our results
are given in Table~\ref{result}. It is found that the $\Lambda_b$ baryon has large decay rates into
both $|\Lambda_c~2^2P_{\lambda} \frac{1}{2}^- \rangle\pi^-$ and $|\Lambda_c~2^2P_{\lambda} \frac{3}{2}^- \rangle\pi^-$ channels.
The branching fraction ratios relative to $\Lambda_c\pi^-$ can reach up to
\begin{eqnarray}\label{1p2p}
\frac{\mathcal{B}\left[\Lambda_b \to \left|\Lambda_c~2^2P_{\lambda} \frac{1}{2}^- \right\rangle \pi^-\right]}{\mathcal{B}[\Lambda_b \rightarrow \Lambda_c^+ \pi^-]} \simeq0.42^{-0.18}_{+0.40},\\
\frac{\mathcal{B}\left[\Lambda_b \to \left|\Lambda_c~2^2P_{\lambda} \frac{3}{2}^- \right\rangle \pi^-\right]}{\mathcal{B}[\Lambda_b \rightarrow \Lambda_c^+ \pi^-]} \simeq0.69^{-0.31}_{+0.71},
\end{eqnarray}
which are about  $3-4$ times larger than the central values predicted in Ref.~\cite{Chua:2019yqh}.
There have been few studies of the $\Lambda_b$ decaying into the higher $2P$-wave $\Lambda_c$ states before.
More theoretical studies are expected to be carried out in the future to better understand the
$\Lambda_b \rightarrow \Lambda_c(2P_{\lambda}) h^-$ processes.

The $\Lambda_b \rightarrow |\Lambda_c~2^2P_{\lambda} \frac{3}{2}^- \rangle \pi^-$ process may have been observed in experiments if the $\Lambda_c(2940)^+$ resonance is assigned to the $|\Lambda_c~2^2P_{\lambda} \frac{3}{2}^- \rangle$ as suggested in the literature~\cite{Lu:2018utx,Yang:2023fsc,Azizi:2022dpn,Gong:2021jkb}.
In 2017, the $\Lambda_b \to \Lambda_c(2940)^+\pi^-\to D^0p\pi^-$ was observed by the LHCb collaboration~\cite{LHCb:2017jym}.
It is found that the spin-parity quantum numbers of $\Lambda_c(2940)^+$ tend to be $J^P=3/2^-$. Furthermore, the collaboration extracted the following branching fraction ratio~\cite{LHCb:2017jym},
\begin{eqnarray}\label{2p12}
\frac{\mathcal{B}[\Lambda_b \to \Lambda_c(2940)^+ \pi^-]\cdot \mathcal{B}[\Lambda_c(2940)^+\to D^0p]}{\mathcal{B}[\Lambda_b \to \Lambda_c(2880)^+ \pi^-]\cdot \mathcal{B}[\Lambda_c(2880)^+\to D^0p]}\simeq 0.83^{+0.31}_{-0.10}\pm 0.06.
\end{eqnarray}
Considering $\Lambda_c(2940)$ as the $|\Lambda_c~2^2P_{\lambda} \frac{3}{2}^- \rangle$ assignment, the branching fraction into the $D^0p$ channel is estimated to be $\mathcal{B}[\Lambda_c(2940)^+\to D^0p]\simeq 40\%$~\cite{Zhang:2024afw,Lu:2018utx}. Together with $\mathcal{B}[\Lambda_c(2880)^+\to D^0p]\simeq 6.3\%$ extracted in Sec.~\ref{1p1a}, from Eq.~(\ref{2p12}) one can extract the ratio $\frac{\mathcal{B}[\Lambda_b \to \Lambda_c(2940)^+ \pi^-]}{\mathcal{B}[\Lambda_b \to \Lambda_c(2880)^+ \pi^-]}$, the central value, $\sim0.13$, is about one order of magnitude smaller than our quark model prediction, $\sim0.92^{-0.10}_{+0.14}$. Thus, we will face a challenge if the $\Lambda_c(2940)^+$ is assigned as a $2P$-wave $\Lambda_c$ state. It should be mentioned that the mass of the $\Lambda_c(2940)^+$ is about $60-100$ MeV lower than the predictions of various
quark models~\cite{Capstick:1986ter,Ebert:2011kk,Chen:2014nyo,Lu:2016ctt}, which indicates it may not be a
conventional baryon, but rather a molecular state as suggested in the literature~\cite{He:2006is,Liang:2011zza,Garcia-Recio:2008rjt, He:2010zq,
Dong:2009tg, Dong:2010xv,Dong:2011ys,Dong:2014ksa, Ortega:2012cx, Ortega:2013fta,Ortega:2014eoa,Ozdem:2023eyz,Zhang:2012jk,
Zhao:2016zhf,Wang:2020dhf, Xie:2015zga,Zhang:2019vqe,Yan:2022nxp,Yan:2023ttx,Yue:2024paz,Yang:2015eoa,Xin:2023gkf}.
%However, when unquenched effects are considered, it remains plausible to classify the $\Lambda_c(2940)^+$ as a $2P$ charmed baryon state~\cite{Cheng:2006dk,Chen:2007xf,Ebert:2007nw,Zhong:2007gp,Liu:2009zg,
%Klempt:2009pi,Chen:2009tm,Lu:2016ctt,Lu:2018utx,Guo:2019ytq,Luo:2019qkm,
%Gandhi:2019xfw,Gong:2021jkb,Yu:2022ymb,Azizi:2022dpn,Zhang:2022pxc,Yu:2023bxn,Yang:2023fsc}.
%It is worth noting that the application of a simple harmonic oscillator to describe the $\Lambda_c(2940)^+$ state introduces certain deviations.
%More reliable studies of the ratio $\frac{\mathcal{B}[\Lambda_b \to \Lambda_c(2940)^+ \pi^-]}{\mathcal{B}[\Lambda_b \to \Lambda_c(2880)^+ \pi^-]}$ are expected to be carried out in theory.

The $\Lambda_c(2910)^+$ resonance recently observed in the $\Sigma_c(2455)^{0,++}\pi^{\pm}$ channels at Belle~\cite{Belle:2022hnm} may be assigned to $|\Lambda_c~2^2P_{\lambda} \frac{1}{2}^- \rangle$ as suggested in the literature~\cite{Yang:2023fsc,Azizi:2022dpn,Weng:2024roa,Luo:2025sns}.
With this assignment, the $\Lambda_c(2910)^+$ resonance should have a large decay rate into $D^0p$ with a branching fraction of $\mathcal{B}[\Lambda_c(2910)^+\to D^0p]\simeq 38\%$~\cite{Zhang:2024afw,Lu:2018utx}. Considering the $\Lambda_c(2910,2940)^+$
as the $2P$-wave doublet, we predict that
\begin{eqnarray}\label{2p32}
\frac{\mathcal{B}[\Lambda_b \to \Lambda_c(2910)^+ \pi^-]\cdot \mathcal{B}[\Lambda_c(2910)^+\to D^0p]}{\mathcal{B}[\Lambda_b \to \Lambda_c(2940)^+ \pi^-]\cdot \mathcal{B}[\Lambda_c(2940)^+\to D^0p]}\simeq 0.58^{+0.02}_{-0.02}.
\end{eqnarray}
This fairly large ratio indicates that with the accumulation of more $\Lambda_b$ samples, the LHCb collaboration should establish the $\Lambda_c(2910)^+$ in the $\Lambda_b \to \Lambda_c(2910)^+\pi^-\to D^0p\pi^-$ decay chain  in forthcoming experiments as well.

To finally establish the $\lambda$-mode $2P$-wave states and clarify the nature of the $\Lambda_c(2910)^+$ and $\Lambda_c(2940)^+$ resonances,
more accurate observations of the $\Lambda_b \to \Lambda_c(2910,2940)^+ \pi^-\to D^0p \pi^-$ processes and the measurements of the relative fraction ratios $\frac{\mathcal{B}[\Lambda_b \to \Lambda_c(2910,2940)^+ \pi^-]}{\mathcal{B}[\Lambda_b \to \Lambda_c^+ \pi^-]}$ are expected to be carried out in future experiments.

\subsection{$\Lambda_b \rightarrow \Lambda_c(2S,3S) h^-$ }

The $\lambda$-mode $2S$ and $3S$ states, denoted by $|\Lambda_c ~~2^2S_{\lambda} \frac{1}{2}^+ \rangle$
and $|\Lambda_c ~~3^2S_{\lambda} \frac{1}{2}^+ \rangle$, belong to the first and second radial excitations
of $\Lambda_c$, respectively. The masses of the $2S$ and $3S$ states are predicted to be $\sim2.8$ and 3.1 GeV, respectively, in quark models~\cite{Ebert:2011kk,Capstick:1986ter,Roberts:2007ni}. In this work, we calculate the branching fraction ratios of the $\Lambda_b \to |\Lambda_c ~2^2S_{\lambda} \frac{1}{2}^+ \rangle h^-$, $|\Lambda_c ~3^2S_{\lambda} \frac{1}{2}^+ \rangle h^-$, processes, and the results are given in Table~\ref{result}.

The $\Lambda_c(2765)^+$ observed in the $\Lambda_c^+\pi^-\pi^+$ final state by the CLEO collaboration may be a good candidate of the $|\Lambda_c ~~2^2S_{\lambda} \frac{1}{2}^+ \rangle$ as suggested in the literature~\cite{Cheng:2006dk,Li:2025frt,Yu:2022ymb,Weng:2024roa}. Considering the $\Lambda_c(2765)^+$ as the $|\Lambda_c ~~2^2S_{\lambda} \frac{1}{2}^+ \rangle$ state, the branching ratio between $\Lambda_c(2765)^+ \pi^-$ and $\Lambda_c^+\pi^-$ channels is predicted to be a fairly large value,
\begin{eqnarray}
\frac{\mathcal{B}[\Lambda_b \rightarrow \Lambda_c(2765)^+ \pi^-]}{\mathcal{B}[\Lambda_b \rightarrow \Lambda_c^+ \pi^-]}\simeq1.01^{-0.33}_{+0.60}.
\end{eqnarray}
It should be mentioned that this branching ratio predicted in this work is about 3 times larger than that of the light-front quark model approach~\cite{Chua:2019yqh}. According to our prediction, the $\Lambda_c(2765)^+$, as a radial excitation of $\Lambda_c^+$, should have a large potential to be established by using the decay chain $\Lambda_b \to \Lambda_c(2765)^+ \pi^-\to (\Sigma_c^{(*)++,0}\pi^{-,+})\pi^-\to (\Lambda_c^+\pi^+\pi^-) \pi^-$.

The high-lying radial excitation $|\Lambda_c ~3^2S_{\lambda} \frac{1}{2}^+ \rangle$ is still missing. It is found that there is a sizeable production rate for this $3S$ state via the pionic weak decay of the $\Lambda_b$ baryon. Taking the predicted mass $M=3130$ MeV~\cite{Ebert:2011kk} for $|\Lambda_c ~3^2S_{\lambda} \frac{1}{2}^+ \rangle$, we have
\begin{eqnarray}
\frac{\mathcal{B}[\Lambda_b \to |\Lambda_c ~~3^2S_{\lambda} \frac{1}{2}^+ \rangle \pi^-]}{\mathcal{B}[\Lambda_b \rightarrow \Lambda_c^+ \pi^-]}\simeq0.19^{-0.09}_{+0.22}.
\end{eqnarray}
The production rate of the $3S$ state is about about 5 times smaller than that of the $2S$ state.
According to quark model predictions, the $|\Lambda_c ~3^2S_{\lambda} \frac{1}{2}^+ \rangle$ dominantly decays into the $D^{*0}p$ and $D^{*+}n$ channels with a comparable branching fraction, $\sim 45\%$. To search for the $|\Lambda_c ~3^2S_{\lambda} \frac{1}{2}^+ \rangle$ state,
the $\Lambda_b \to \Lambda_c (3S) \pi^-\to D^{*0}p \pi^-$ decay chain is worth observing in future experiments.

\subsection{$\Lambda_b \rightarrow \Lambda_c(2D,1F) h^-$ }

There are two $\lambda$-mode $2D$-wave states, $|\Lambda_c~2^2D_{\lambda} \frac{3}{2}^+ \rangle$ and $|\Lambda_c~2^2D_{\lambda} \frac{5}{2}^+ \rangle$, and two $\lambda$-mode $1F$-wave states, $|\Lambda_c~1^2F_{\lambda} \frac{5}{2}^- \rangle$ and $|\Lambda_c~1^2F_{\lambda} \frac{7}{2}^- \rangle$ according to quark model classification. These states are still missing. In theory, the masses for the $1F$ and $2D$-wave $\Lambda_c$ states are predicted to be $\sim3.1$ and $\sim3.2$ GeV, respectively~\cite{Lu:2019rtg,Yu:2022ymb,Luo:2019qkm,Azizi:2022dpn,Yang:2023fsc,
Roberts:2007ni,Capstick:1986ter,Ebert:2011kk}. In this work,  using the predicted masses from Ref.~\cite{Ebert:2011kk} we calculate the branching fraction ratios of the $\Lambda_b \rightarrow \Lambda_c(2D_\lambda,1F_\lambda) h^-$ ($h^-=\pi^-, \rho^-, K^{-}, K^{\ast-}, D^{-}, D^{\ast-}, D_s^{-}, D_s^{\ast-}$) processes for the first time, and our results are given in Table~\ref{result}.

It is found that both the $1F$ and $2D$-wave $\Lambda_c$ states have significant production rates via the pionic nonleptonic weak decays of $\Lambda_b$. The branching ratios relative to $\Lambda_c^+\pi$ are predicted to be
\begin{eqnarray}\label{1p2p}
\frac{\mathcal{B}\left[\Lambda_b \to \left|\Lambda_c~2^2D_{\lambda} \frac{3}{2}^+ \right\rangle \pi^-\right]}{\mathcal{B}[\Lambda_b \rightarrow \Lambda_c^+ \pi^-]} \simeq0.19^{-0.10}_{+0.27},~\frac{\mathcal{B}\left[\Lambda_b \to \left|\Lambda_c~2^2D_{\lambda} \frac{5}{2}^+ \right\rangle \pi^-\right]}{\mathcal{B}[\Lambda_b \rightarrow \Lambda_c^+ \pi^-]} \simeq0.20^{-0.11}_{+0.31},\\
\frac{\mathcal{B}\left[\Lambda_b \to \left|\Lambda_c~1^2F_{\lambda} \frac{5}{2}^- \right\rangle \pi^-\right]}{\mathcal{B}[\Lambda_b \rightarrow \Lambda_c^+ \pi^-]} \simeq0.19^{-0.09}_{+0.22},~\frac{\mathcal{B}\left[\Lambda_b \to \left|\Lambda_c~1^2F_{\lambda} \frac{7}{2}^- \right\rangle \pi^-\right]}{\mathcal{B}[\Lambda_b \rightarrow \Lambda_c^+ \pi^-]} \simeq0.26^{-0.12}_{+0.30}.
\end{eqnarray}
According to the predictions of the OZI-allowed strong decays of the excited $\Lambda_c$ states in Ref.~\cite{Lu:2019rtg},
the $|\Lambda_c~2^2D_{\lambda} \frac{5}{2}^+ \rangle$ and $|\Lambda_c~1^2F_{\lambda} \frac{5}{2}^- \rangle$ states have significant decay rates into the $D^0p$ channel with a branching fraction of $\sim 20-30\%$. While the $|\Lambda_c~2^2D_{\lambda} \frac{3}{2}^+ \rangle$ and $|\Lambda_c~1^2F_{\lambda} \frac{7}{2}^- \rangle$ states have significant decay rates into the $D^{*0}p$ channel with a branching fraction of $\sim 40-50\%$. Thus, the $1F$ and $2D$-wave $\Lambda_c$ states are likely to be established in the decays $\Lambda_b\to \Lambda_c(2D,1F) \pi^-\to D^0p \pi^-, D^{*0}p\pi^-$.

\section{Summary}\label{SUM}

In this work, we systematically study the production of the $\Lambda_c$ baryon and its excitations via the nonleptonic weak decays of the $\Lambda_b$ baryon within a constituent quark model.
The existing branching fractions measured in experiments for the $\Lambda_b \rightarrow \Lambda_c(\pi^-, K^-, D^-, D_s^-,D_s^{*-})$ processes
can be well described by theoretical calculations.

Furthermore, we present our predictions of the production rates of the $2S$-, $3S$-, $1P$-, $2P$-, $1D$-, $2D$-, and $1F$-wave $\Lambda_c^+$ excitations in the $\Lambda_b \rightarrow \Lambda_c^{*}(\pi^-, \rho^-, K^{-}, K^{\ast-}, D^{-}, D^{\ast-}, D_s^{-}, D_s^{\ast-})$ processes. It is found that the excited $\Lambda_c$ baryons have a large production rate in the $\Lambda_b$ nonleptonic weak decay process associated with  $\pi^-$ meson emission. Considering the $\Lambda_c(2595,2625)^+$ as the $1P$-wave $\Lambda_c^+$ states,
while $\Lambda_c(2860,2880)^+$ as the $1D$-wave states, the observations of the $\Lambda_b \to \Lambda_c(2595,2625)^+\pi^-\to \Lambda_c^+ \pi^- \pi^+\pi^-$ and $\Lambda_b \to \Lambda_c(2860,2880)^+\pi^-\to D^0p\pi^-$ decays are consistent with our quark model predictions as well. However, if we assign the $\Lambda_c(2940)^+$ as a $2P$-wave $\Lambda_c$ state, we cannot explain the $\Lambda_b \to \Lambda_c(2940)^+\pi^-\to D^0p\pi^-$ decay rate observed at LHCb.

To establish the $2P$-wave $\Lambda_c$ states and clarify the nature of the $\Lambda_c(2910)^+$ and $\Lambda_c(2940)^+$ resonances,
more accurate observations of the $\Lambda_b \to \Lambda_c(2910,2940)^+ \pi^-\to D^0p \pi^-$ processes are expected to be carried out in future experiments.

To establish the $2S$-wave $\Lambda_c$ state and clarify the nature of the $\Lambda_c(2765)^+$, the decay chain $\Lambda_b \to \Lambda_c(2765)^+ \pi^-\to (\Sigma_c^{(*)++,0}\pi^{-,+})\pi^-\to (\Lambda_c^+\pi^+\pi^-) \pi^-$ is expected to be observed in future experiments.

To search for the high-lying $3S$-, $1F$-, and $2D$-wave $\Lambda_c$ states,
the $\Lambda_b \to \Lambda_c (3S) \pi^-\to D^{*0}p \pi^-$, and $\Lambda_b\to \Lambda_c(2D,1F) \pi^-\to D^0p \pi^-, D^{*0}p\pi^-$ decays are worth observing in future experiments.

Finally, it should be pointed out that there is a notable model dependency in the predictions of the $\Lambda_b \to \Lambda_c(\rho^-, K^{*-})$ decays. These processes are also expected to be observed in future experiments to test different model predictions.

\section*{Acknowledgements }

We would like to thank  Yu-Kuo Hsiao for very helpful discussions.
This work is supported by the National Natural Science Foundation of
China (Grants No.12205026, No.12235018, and No.12175065), and Applied Basic Research Program of Shanxi Province,
China under Grant No.202103021223376 and 202203021212171, and Support Project for Young Teachers' Research
and Innovation Abilities under Grant No.2025Q037.

%\bibliographystyle{unsrt}
%\bibliography{wkl}

\begin{thebibliography}{10}


%1\cite{Knapp:1976qw}
\bibitem{Knapp:1976qw}
B.~Knapp, W.~Y.~Lee, P.~Leung, S.~D.~Smith, A.~Wijangco, J.~Knauer, D.~Yount, J.~Bronstein, R.~Coleman and G.~Gladding, \textit{et al.}
Observation of a Narrow anti-Baryon State at 2.26-GeV/c$^2$,
Phys. Rev. Lett. \textbf{37}, 882 (1976).
%doi:10.1103/PhysRevLett.37.882
%376 citations counted in INSPIRE as of 25 Nov 2025

%2\cite{Chen:2022asf}
\bibitem{Chen:2022asf}
H.~X.~Chen, W.~Chen, X.~Liu, Y.~R.~Liu and S.~L.~Zhu,
An updated review of the new hadron states,
Rept. Prog. Phys. \textbf{86},  026201 (2023).
%doi:10.1088/1361-6633/aca3b6
%[arXiv:2204.02649 [hep-ph]].
%629 citations counted in INSPIRE as of 25 Nov 2025

%3\cite{Cheng:2021qpd}
\bibitem{Cheng:2021qpd}
H.~Y.~Cheng,
Charmed baryon physics circa 2021,
Chin. J. Phys. \textbf{78}, 324-362 (2022).
%doi:10.1016/j.cjph.2022.06.021
%[arXiv:2109.01216 [hep-ph]].
%111 citations counted in INSPIRE as of 25 Nov 2025





%4\cite{Chen:2016spr}
\bibitem{Chen:2016spr}
H.~X.~Chen, W.~Chen, X.~Liu, Y.~R.~Liu and S.~L.~Zhu,
A review of the open charm and open bottom systems,
Rept. Prog. Phys. \textbf{80},  076201 (2017).
%doi:10.1088/1361-6633/aa6420
%[arXiv:1609.08928 [hep-ph]].
%441 citations counted in INSPIRE as of 25 Nov 2025

%5\cite{Eichmann:2016yit}
\bibitem{Eichmann:2016yit}
G.~Eichmann, H.~Sanchis-Alepuz, R.~Williams, R.~Alkofer and C.~S.~Fischer,
Baryons as relativistic three-quark bound states,
Prog. Part. Nucl. Phys. \textbf{91}, 1-100 (2016).
%doi:10.1016/j.ppnp.2016.07.001
%[arXiv:1606.09602 [hep-ph]].
%455 citations counted in INSPIRE as of 25 Nov 2025



%6\cite{Cheng:2015iom}
\bibitem{Cheng:2015iom}
H.~Y.~Cheng,
Charmed baryons circa 2015,
Front. Phys. (Beijing) \textbf{10}, 101406 (2015).
%doi:10.1007/s11467-015-0483-z
%109 citations counted in INSPIRE as of 25 Nov 2025




%7\cite{Crede:2013kia}
\bibitem{Crede:2013kia}
V.~Crede and W.~Roberts,
Progress towards understanding baryon resonances,
Rept. Prog. Phys. \textbf{76}, 076301 (2013).
%doi:10.1088/0034-4885/76/7/076301
%[arXiv:1302.7299 [nucl-ex]].
%296 citations counted in INSPIRE as of 25 Nov 2025



%8\cite{Klempt:2009pi}
\bibitem{Klempt:2009pi}
E.~Klempt and J.~M.~Richard,
Baryon spectroscopy,
Rev. Mod. Phys. \textbf{82}, 1095-1153 (2010).
%doi:10.1103/RevModPhys.82.1095
%[arXiv:0901.2055 [hep-ph]].
%447 citations counted in INSPIRE as of 25 Nov 2025








%9\cite{Meng:2022ozq}
\bibitem{Meng:2022ozq}
L.~Meng, B.~Wang, G.~J.~Wang and S.~L.~Zhu,
Chiral perturbation theory for heavy hadrons and chiral effective field theory for heavy hadronic molecules,
Phys. Rept. \textbf{1019}, 1-149 (2023).
%doi:10.1016/j.physrep.2023.04.003
%[arXiv:2204.08716 [hep-ph]].
%263 citations counted in INSPIRE as of 25 Nov 2025

%10\cite{ParticleDataGroup:2024cfk}
\bibitem{pdg}
S.~Navas \textit{et al.} [Particle Data Group],
Review of particle physics,
Phys. Rev. D \textbf{110},  030001 (2024).
%doi:10.1103/PhysRevD.110.030001
%3254 citations counted in INSPIRE as of 25 Nov 2025

%11\cite{Capstick:1986ter}
\bibitem{Capstick:1986ter}
S.~Capstick and N.~Isgur,
Baryons in a relativized quark model with chromodynamics,
Phys. Rev. D \textbf{34},  2809-2835 (1986).
%doi:10.1103/physrevd.34.2809
%1552 citations counted in INSPIRE as of 25 Nov 2025

%12\cite{Cheng:2006dk}
\bibitem{Cheng:2006dk}
H.~Y.~Cheng and C.~K.~Chua,
Strong Decays of Charmed Baryons in Heavy Hadron Chiral Perturbation Theory,
Phys. Rev. D \textbf{75}, 014006 (2007).
%doi:10.1103/PhysRevD.75.014006
%[arXiv:hep-ph/0610283 [hep-ph]].
%163 citations counted in INSPIRE as of 25 Nov 2025

%13\cite{Cheng:2015naa}
\bibitem{Cheng:2015naa}
H.~Y.~Cheng and C.~K.~Chua,
Strong Decays of Charmed Baryons in Heavy Hadron Chiral Perturbation Theory: An Update,
Phys. Rev. D \textbf{92}, 074014 (2015).
%doi:10.1103/PhysRevD.92.074014
%[arXiv:1508.05653 [hep-ph]].
%98 citations counted in INSPIRE as of 25 Nov 2025

%14\cite{Chen:2007xf}
\bibitem{Chen:2007xf}
C.~Chen, X.~L.~Chen, X.~Liu, W.~Z.~Deng and S.~L.~Zhu,
Strong decays of charmed baryons,
Phys. Rev. D \textbf{75}, 094017 (2007).
%doi:10.1103/PhysRevD.75.094017
%[arXiv:0704.0075 [hep-ph]].
%152 citations counted in INSPIRE as of 25 Nov 2025

%15\cite{Chen:2015kpa}
\bibitem{Chen:2015kpa}
H.~X.~Chen, W.~Chen, Q.~Mao, A.~Hosaka, X.~Liu and S.~L.~Zhu,
$P$-wave charmed baryons from QCD sum rules,
Phys. Rev. D \textbf{91}, 054034 (2015).
%doi:10.1103/PhysRevD.91.054034
%[arXiv:1502.01103 [hep-ph]].
%118 citations counted in INSPIRE as of 25 Nov 2025

%16\cite{Chen:2016phw}
\bibitem{Chen:2016phw}
H.~X.~Chen, Q.~Mao, A.~Hosaka, X.~Liu and S.~L.~Zhu,
$D$-wave charmed and bottomed baryons from QCD sum rules,
Phys. Rev. D \textbf{94}, 114016 (2016).
%doi:10.1103/PhysRevD.94.114016
%[arXiv:1611.02677 [hep-ph]].
%83 citations counted in INSPIRE as of 25 Nov 2025

%17\cite{Chen:2017sci}
\bibitem{Chen:2017sci}
H.~X.~Chen, Q.~Mao, W.~Chen, A.~Hosaka, X.~Liu and S.~L.~Zhu,
Decay properties of $P$-wave charmed baryons from light-cone QCD sum rules,
Phys. Rev. D \textbf{95}, 094008 (2017).
%doi:10.1103/PhysRevD.95.094008
%[arXiv:1703.07703 [hep-ph]].
%118 citations counted in INSPIRE as of 25 Nov 2025

%18\cite{Ebert:2007nw}
\bibitem{Ebert:2007nw}
D.~Ebert, R.~N.~Faustov and V.~O.~Galkin,
Masses of excited heavy baryons in the relativistic quark model,
Phys. Lett. B \textbf{659}, 612-620 (2008).
%doi:10.1016/j.physletb.2007.11.037
%[arXiv:0705.2957 [hep-ph]].
%267 citations counted in INSPIRE as of 25 Nov 2025

%19\cite{Ebert:2011kk}
\bibitem{Ebert:2011kk}
D.~Ebert, R.~N.~Faustov and V.~O.~Galkin,
Spectroscopy and Regge trajectories of heavy baryons in the relativistic quark-diquark picture,
Phys. Rev. D \textbf{84}, 014025 (2011).
%doi:10.1103/PhysRevD.84.014025
%[arXiv:1105.0583 [hep-ph]].
%386 citations counted in INSPIRE as of 25 Nov 2025

%20\cite{Roberts:2007ni}
\bibitem{Roberts:2007ni}
W.~Roberts and M.~Pervin,
Heavy baryons in a quark model,
Int. J. Mod. Phys. A \textbf{23}, 2817-2860 (2008).
%doi:10.1142/S0217751X08041219
%[arXiv:0711.2492 [nucl-th]].
%529 citations counted in INSPIRE as of 25 Nov 2025

%21\cite{Chen:2009tm}
\bibitem{Chen:2009tm}
B.~Chen, D.~X.~Wang and A.~Zhang,
$J^P$ Assignments of $\Lambda_c^+$ Baryons,''
Chin. Phys. C \textbf{33}, 1327-1330 (2009).
%doi:10.1088/1674-1137/33/12/047
%[arXiv:0906.3934 [hep-ph]].
%39 citations counted in INSPIRE as of 25 Nov 2025

%22\cite{Chen:2014nyo}
\bibitem{Chen:2014nyo}
B.~Chen, K.~W.~Wei and A.~Zhang,
Assignments of $\Lambda_Q$ and $\Xi_Q$ baryons in the heavy quark-light diquark picture,
Eur. Phys. J. A \textbf{51}, 82 (2015).
%doi:10.1140/epja/i2015-15082-3
%[arXiv:1406.6561 [hep-ph]].
%151 citations counted in INSPIRE as of 25 Nov 2025

%23\cite{Chen:2016iyi}
\bibitem{Chen:2016iyi}
B.~Chen, K.~W.~Wei, X.~Liu and T.~Matsuki,
Low-lying charmed and charmed-strange baryon states,
Eur. Phys. J. C \textbf{77}, 154 (2017).
%doi:10.1140/epjc/s10052-017-4708-x
%[arXiv:1609.07967 [hep-ph]].
%106 citations counted in INSPIRE as of 25 Nov 2025

%24\cite{Zhong:2007gp}
\bibitem{Zhong:2007gp}
X.~H.~Zhong and Q.~Zhao,
Charmed baryon strong decays in a chiral quark model,
Phys. Rev. D \textbf{77}, 074008 (2008).


%25\cite{Wang:2017kfr}
\bibitem{Wang:2017kfr}
K.~L.~Wang, Y.~X.~Yao, X.~H.~Zhong and Q.~Zhao,
Strong and radiative decays of the low-lying $S$- and $P$-wave singly heavy baryons,
Phys. Rev. D \textbf{96}, 116016 (2017).


%26\cite{Yoshida:2015tia}
\bibitem{Yoshida:2015tia}
T.~Yoshida, E.~Hiyama, A.~Hosaka, M.~Oka and K.~Sadato,
Spectrum of heavy baryons in the quark model,
Phys. Rev. D \textbf{92}, 114029 (2015).

%26\cite{Xie:2025gom}
\bibitem{Xie:2025gom}
X.~Xie, J.~Tong, Q.~Huang, H.~Huang and J.~Ping,
Strong decays of singly heavy baryons,
[arXiv:2504.10183 [hep-ph]].
%1 citations counted in INSPIRE as of 25 Nov 2025


%27\cite{Zhang:2024afw}
\bibitem{Zhang:2024afw}
Y.~B.~Zhang, L.~Y.~Xiao and X.~H.~Zhong,
Possible explanations of the observed $\Lambda_c$ resonances,
Chin. Phys. \textbf{49}, 112001 (2025).

%28\cite{Yu:2023bxn}
\bibitem{Yu:2023bxn}
G.~L.~Yu, Y.~Meng, Z.~Y.~Li, Z.~G.~Wang and L.~Jie,
Strong decay properties of single heavy baryons $\Lambda_Q$, $\Sigma_Q$ and $\Omega_Q$,
Int. J. Mod. Phys. A \textbf{38}, 2350082 (2023).

%29\cite{Shah:2016mig}
\bibitem{Shah:2016mig}
Z.~Shah, K.~Thakkar, A.~Kumar Rai and P.~C.~Vinodkumar,
Excited State Mass spectra of Singly Charmed Baryons,
Eur. Phys. J. A \textbf{52}, 313 (2016).

%30\cite{Tawfiq:1998nk}
\bibitem{Tawfiq:1998nk}
S.~Tawfiq, P.~J.~O'Donnell and J.~G.~Korner,
Charmed baryon strong coupling constants in a light front quark model,
Phys. Rev. D \textbf{58}, 054010 (1998).

%31\cite{Ivanov:1999bk}
\bibitem{Ivanov:1999bk}
M.~A.~Ivanov, J.~G.~Korner, V.~E.~Lyubovitskij and A.~G.~Rusetsky,
Strong and radiative decays of heavy flavored baryons,
Phys. Rev. D \textbf{60}, 094002 (1999).




%32\cite{Chow:1995nw}
\bibitem{Chow:1995nw}
C.~K.~Chow,
Radiative decays of excited $\Lambda_Q$ baryons in the bound state picture,
Phys. Rev. D \textbf{54}, 3374-3376 (1996).

%33\cite{Pirjol:1997nh}
\bibitem{Pirjol:1997nh}
D.~Pirjol and T.~M.~Yan,
Predictions for $s$ wave and $p$ wave heavy baryons from sum rules and constituent quark model. 1. Strong interactions,
Phys. Rev. D \textbf{56}, 5483-5510 (1997).

%34\cite{Lin:2021wrb}
\bibitem{Lin:2021wrb}
Q.~Y.~Lin and X.~Liu,
Production of charmed baryon $\Lambda_c(2860)$ via low energy antiproton-proton interaction,
Phys. Rev. D \textbf{105}, 014035 (2022).

%35\cite{Kim:2020imk}
\bibitem{Kim:2020imk}
Y.~Kim, E.~Hiyama, M.~Oka and K.~Suzuki,
Spectrum of singly heavy baryons from a chiral effective theory of diquarks,
Phys. Rev. D \textbf{102}, 014004 (2020).

%36\cite{Kim:2024tbf}
\bibitem{Kim:2024tbf}
Y.~Kim, M.~Oka and K.~Suzuki,
Chiral effective theory of scalar and vector diquarks revisited,
Phys. Rev. D \textbf{111}, 034014 (2025).

%37\cite{Ponkhuha:2024gms}
\bibitem{Ponkhuha:2024gms}
N.~Ponkhuha, A.~J.~Arifi and D.~Samart,
Two-pion emission decays of negative parity singly heavy baryons,
Phys. Rev. D \textbf{110}, 114046 (2024).

%38\cite{Arifi:2021orx}
\bibitem{Arifi:2021orx}
A.~J.~Arifi, D.~Suenaga and A.~Hosaka,
Relativistic corrections to decays of heavy baryons in the quark model,
Phys. Rev. D \textbf{103}, 094003 (2021).


%39\cite{Garcia-Tecocoatzi:2022zrf}
\bibitem{Garcia-Tecocoatzi:2022zrf}
H.~Garcia-Tecocoatzi, A.~Giachino, J.~Li, A.~Ramirez-Morales and E.~Santopinto,
Strong decay widths and mass spectra of charmed baryons,
Phys. Rev. D \textbf{107},  034031 (2023).

%40\cite{Gong:2021jkb}
\bibitem{Gong:2021jkb}
K.~Gong, H.~Y.~Jing and A.~Zhang,
Possible assignments of highly excited $\Lambda _c(2860)^+$, $\Lambda _c(2880)^+$ and $\Lambda _c(2940)^+$,
Eur. Phys. J. C \textbf{81}, 467 (2021).


%41\cite{Yao:2018jmc}
\bibitem{Yao:2018jmc}
Y.~X.~Yao, K.~L.~Wang and X.~H.~Zhong,
Strong and radiative decays of the low-lying $D$-wave singly heavy baryons,
Phys. Rev. D \textbf{98}, 076015 (2018).



%42\cite{Chen:2017aqm}
\bibitem{Chen:2017aqm}
B.~Chen, X.~Liu and A.~Zhang,
Newly observed $\Lambda_c(2860)^+$ at LHCb and its $D$-wave partners $\Lambda_c(2880)^+$, $\Xi_c(3055)^+$ and $\Xi_c(3080)^+$,
Phys. Rev. D \textbf{95}, 074022 (2017).


%43\cite{Guo:2019ytq}
\bibitem{Guo:2019ytq}
J.~J.~Guo, P.~Yang and A.~Zhang,
Strong decays of observed $\Lambda_c$ baryons in the $^3P_0$ model,
Phys. Rev. D \textbf{100}, 014001 (2019).


%44\cite{Lu:2019rtg}
\bibitem{Lu:2019rtg}
Q.~F.~L{\"u} and X.~H.~Zhong,
Strong decays of the higher excited $\Lambda_Q$ and $\Sigma_Q$ baryons,
Phys. Rev. D \textbf{101}, 014017 (2020).


%45\cite{Lu:2018utx}
\bibitem{Lu:2018utx}
Q.~F.~L{\"u}, L.~Y.~Xiao, Z.~Y.~Wang and X.~H.~Zhong,
Strong decay of $\Lambda _c(2940)$ as a $2P$ state in the $\Lambda _c$ family,
Eur. Phys. J. C \textbf{78}, 599 (2018).


%46\cite{Yang:2023fsc}
\bibitem{Yang:2023fsc}
H.~M.~Yang and H.~X.~Chen,
$2P$-wave charmed baryons from QCD sum rules,
Phys. Rev. D \textbf{109}, 036032 (2024).


%47\cite{Azizi:2022dpn}
\bibitem{Azizi:2022dpn}
K.~Azizi, Y.~Sarac and H.~Sundu,
Interpretation of the $\Lambda _c(2910)^+$ baryon newly seen by Belle Collaboration and its possible bottom partner,
Eur. Phys. J. C \textbf{82}, 920 (2022).


%48\cite{Weng:2024roa}
\bibitem{Weng:2024roa}
X.~Z.~Weng, W.~Z.~Deng and S.~L.~Zhu,
Heavy baryons in the relativized quark model with chromodynamics,
Phys. Rev. D \textbf{110}, 056052 (2024).

%49\cite{Lu:2016ctt}
\bibitem{Lu:2016ctt}
Q.~F.~L{\"u}, Y.~Dong, X.~Liu and T.~Matsuki,
Puzzle of the $\Lambda_c$ Spectrum,
Nucl. Phys. Rev. \textbf{35}, 1-4 (2018).


%50\cite{Luo:2019qkm}
\bibitem{Luo:2019qkm}
S.~Q.~Luo, B.~Chen, Z.~W.~Liu and X.~Liu,
Resolving the low mass puzzle of $\Lambda_c(2940)^+$,
Eur. Phys. J. C \textbf{80}, 301 (2020).


%51\cite{Gandhi:2019xfw}
\bibitem{Gandhi:2019xfw}
K.~Gandhi, Z.~Shah and A.~K.~Rai,
Spectrum of nonstrange singly charmed baryons in the constituent quark model,
Int. J. Theor. Phys. \textbf{59}, 1129-1156 (2020).


%52\cite{Yu:2022ymb}
\bibitem{Yu:2022ymb}
G.~L.~Yu, Z.~Y.~Li, Z.~G.~Wang, J.~Lu and M.~Yan,
Systematic analysis of single heavy baryons $\Lambda_Q$, $\Sigma_Q$ and $\Omega_Q$,
Nucl. Phys. B \textbf{990}, 116183 (2023).


%53\cite{Zhang:2022pxc}
\bibitem{Zhang:2022pxc}
Z.~L.~Zhang, Z.~W.~Liu, S.~Q.~Luo, F.~L.~Wang, B.~Wang and H.~Xu,
$\Lambda_c(2910)$ and $\Lambda_c(2940)$ as conventional baryons dressed with the $D^{\ast}N$ channel,
Phys. Rev. D \textbf{107}, 034036 (2023).


%54\cite{Liu:2009zg}
\bibitem{Liu:2009zg}
X.~Liu,
Strong decays of newly observed heavy flavor hadrons,
Chin. Phys. C \textbf{33}, 473-480 (2009).


%55\cite{He:2006is}
\bibitem{He:2006is}
X.~G.~He, X.~Q.~Li, X.~Liu and X.~Q.~Zeng,
$\Lambda^+_c(2940)$: A Possible molecular state?,
Eur. Phys. J. C \textbf{51}, 883-889 (2007).


%56\cite{Garcia-Recio:2008rjt}
\bibitem{Garcia-Recio:2008rjt}
C.~Garcia-Recio, V.~K.~Magas, T.~Mizutani, J.~Nieves, A.~Ramos, L.~L.~Salcedo and L.~Tolos,
The $s$-wave charmed baryon resonances from a coupled-channel approach with heavy quark symmetry,
Phys. Rev. D \textbf{79}, 054004 (2009).


%57\cite{Dong:2009tg}
\bibitem{Dong:2009tg}
Y.~Dong, A.~Faessler, T.~Gutsche and V.~E.~Lyubovitskij,
Strong two-body decays of the $\Lambda_c(2940)^+$ in a hadronic molecule picture,
Phys. Rev. D \textbf{81}, 014006 (2010).


%58\cite{Dong:2010xv}
\bibitem{Dong:2010xv}
Y.~Dong, A.~Faessler, T.~Gutsche, S.~Kumano and V.~E.~Lyubovitskij,
Radiative decay of $\Lambda_c(2940)^+$ in a hadronic molecule picture,
Phys. Rev. D \textbf{82}, 034035 (2010).


%59\cite{He:2010zq}
\bibitem{He:2010zq}
J.~He, Y.~T.~Ye, Z.~F.~Sun and X.~Liu,
The observed charmed hadron $\Lambda_c(2940)^+$ and the $D^*N$ interaction,
Phys. Rev. D \textbf{82}, 114029 (2010).


%60\cite{Liang:2011zza}
\bibitem{Liang:2011zza}
W.~H.~Liang, Y.~F.~Qiu, J.~X.~Hu and P.~N.~Shen,
$\Lambda^*_c$ in a coupled-channel baryon-meson scattering,
Chin. Phys. C \textbf{35}, 16-21 (2011).


%61\cite{Dong:2011ys}
\bibitem{Dong:2011ys}
Y.~Dong, A.~Faessler, T.~Gutsche, S.~Kumano and V.~E.~Lyubovitskij,
Strong three-body decays of $\Lambda_c(2940)^+$,
Phys. Rev. D \textbf{83}, 094005 (2011).


%62\cite{Ortega:2012cx}
\bibitem{Ortega:2012cx}
P.~G.~Ortega, D.~R.~Entem and F.~Fernandez,
Quark model description of the $\Lambda_c(2940)^+$ as a molecular $D^*N$ state and the possible existence of the $\Lambda_b(6248)$,
Phys. Lett. B \textbf{718}, 1381-1384 (2013).




%63\cite{Zhang:2012jk}
\bibitem{Zhang:2012jk}
J.~R.~Zhang,
$S$-wave $D^{(*)}N$ molecular states: $\Sigma_{c}(2800)$ and $\Lambda_{c}(2940)^{+}$?,
Phys. Rev. D \textbf{89}, 096006 (2014).


%64\cite{Ortega:2013fta}
\bibitem{Ortega:2013fta}
P.~G.~Ortega, D.~R.~Entem and F.~Fernandez,
The $\Lambda_{c}(2940)^+$ as a $D^*N$ Molecule in a Constituent Quark Model and a Possible $\Lambda_{b}(6248)$,
Few Body Syst. \textbf{54},  1101-1104 (2013).


%65\cite{Dong:2014ksa}
\bibitem{Dong:2014ksa}
Y.~Dong, A.~Faessler, T.~Gutsche and V.~E.~Lyubovitskij,
Role of the hadron molecule $\Lambda_c$(2940) in the $p\bar{p} \to pD^0\bar{\Lambda}_c$(2286) annihilation reaction,
Phys. Rev. D \textbf{90}, 094001 (2014).


%66\cite{Ortega:2014eoa}
\bibitem{Ortega:2014eoa}
P.~G.~Ortega, D.~R.~Entem and F.~Fern{\'a}ndez,
Hadronic molecules in the open charm and open bottom baryon spectrum,
Phys. Rev. D \textbf{90},  114013 (2014).

%67\cite{Xie:2015zga}
\bibitem{Xie:2015zga}
J.~J.~Xie, Y.~B.~Dong and X.~Cao,
Role of the $\Lambda^+_c(2940)$ in the $\pi^- p \to D^- D^0 p$ reaction close to threshold,
Phys. Rev. D \textbf{92}, 034029 (2015).


%68\cite{Yang:2015eoa}
\bibitem{Yang:2015eoa}
D.~Yang, J.~Liu and D.~Zhang,
$N D^{(*)}$ and $N B^{(*)}$ interactions in a chiral quark model,
[arXiv:1508.03883 [nucl-th]].
%4 citations counted in INSPIRE as of 25 Nov 2025

%69\cite{Zhao:2016zhf}
\bibitem{Zhao:2016zhf}
L.~Zhao, H.~Huang and J.~Ping,
$ND$ and $NB$ systems in quark delocalization color screening model,
Eur. Phys. J. A \textbf{53}, 28 (2017).


%70\cite{Zhang:2019vqe}
\bibitem{Zhang:2019vqe}
D.~Zhang, D.~Yang, X.~F.~Wang and K.~Nakayama,
Possible $S$-wave $ND^{(*)}$ and $N\bar B^{(*)}$ bound states in a chiral quark model,
[arXiv:1903.01207 [nucl-th]].
%9 citations counted in INSPIRE as of 25 Nov 2025

%71\cite{Wang:2020dhf}
\bibitem{Wang:2020dhf}
B.~Wang, L.~Meng and S.~L.~Zhu,
$D^{(\ast)}N$ interaction and the structure of $\Sigma_c(2800)$ and $\Lambda_c(2940)$ in chiral effective field theory,
Phys. Rev. D \textbf{101},  094035 (2020).


%72\cite{Yan:2022nxp}
\bibitem{Yan:2022nxp}
Y.~Yan, X.~Hu, Y.~Wu, H.~Huang, J.~Ping and Y.~Yang,
Pentaquark interpretation of $\Lambda _{c}$ states in the quark model,
Eur. Phys. J. C \textbf{83}, 524 (2023).


%73\cite{Xin:2023gkf}
\bibitem{Xin:2023gkf}
Q.~Xin, X.~S.~Yang and Z.~G.~Wang,
The singly charmed pentaquark molecular states via the QCD sum rules,
Int. J. Mod. Phys. A \textbf{38},  2350123 (2023).


%74\cite{Ozdem:2023eyz}
\bibitem{Ozdem:2023eyz}
U.~Ozdem,
Electromagnetic properties of the $\Sigma_{c}(2800)^+$ and $\Lambda_c(2940)^+$ states via light-cone QCD,
Eur. Phys. J. C \textbf{83}, 1077 (2023).


%75\cite{Yan:2023ttx}
\bibitem{Yan:2023ttx}
M.~J.~Yan, F.~Z.~Peng and M.~Pavon Valderrama,
Molecular charmed baryons and pentaquarks from light-meson exchange saturation,
Phys. Rev. D \textbf{109},  1 (2024).


%76\cite{Yue:2024paz}
\bibitem{Yue:2024paz}
Z.~L.~Yue, Q.~Y.~Guo and D.~Y.~Chen,
Strong decays of the $\Lambda_c(2910)$ and $\Lambda_c(2940)$ in the $ND^{\ast}$ molecular frame,
Phys. Rev. D \textbf{109}, 094049 (2024).


%77\cite{Guo:2025tuz}
\bibitem{Guo:2025tuz}
Q.~Y.~Guo and D.~Y.~Chen,
$\Lambda_{c}(2910)$ and $\Lambda_{c}(2940)$ productions in $p \bar{p}$ annihilation and $K^{-}p$ scattering processes,
[arXiv:2509.21176 [hep-ph]].
%0 citations counted in INSPIRE as of 25 Nov 2025

%78\cite{Wang:2022dmw}
\bibitem{Wang:2022dmw}
W.~J.~Wang, L.~Y.~Xiao and X.~H.~Zhong,
Strong decays of the low-lying $\rho$-mode $1P$-wave singly heavy baryons,
Phys. Rev. D \textbf{106},  074020 (2022).


%79\cite{Wang:2024sbw}
\bibitem{Wang:2024sbw}
F.~L.~Wang, S.~Q.~Luo and X.~Liu,
Unveiling the composition of the single-charm molecular pentaquarks: insights from radiative decay and magnetic moment,
Eur. Phys. J. C \textbf{85}, 216 (2025).


%80\cite{Li:2025frt}
\bibitem{Li:2025frt}
Z.~Y.~Li, G.~L.~Yu, Z.~G.~Wang, J.~Z.~Gu and H.~T.~Shen,
Mass spectra of singly heavy baryons in the relativized quark model with heavy-quark dominance,
Chin. Phys. \textbf{49},  113107 (2025).



%81\cite{LHCb:2017jym}
\bibitem{LHCb:2017jym}
R.~Aaij \textit{et al.} [LHCb],
Study of the $D^0 p$ amplitude in $\Lambda_b^0\to D^0 p \pi^-$ decays,
JHEP \textbf{05}, 030 (2017).


%88\cite{CLEO:2000mbh}
\bibitem{CLEO:2000mbh}
M.~Artuso \textit{et al.} [CLEO],
Observation of new states decaying into $\Lambda^+_c \pi^- \pi^+$,
Phys. Rev. Lett. \textbf{86}, 4479-4482 (2001).


%83\cite{Belle:2006xni}
\bibitem{Belle:2006xni}
K.~Abe \textit{et al.} [Belle],
Experimental constraints on the possible $J^P$ quantum numbers of the $\Lambda_c(2880)^+$,
Phys. Rev. Lett. \textbf{98}, 262001 (2007).


%84\cite{BaBar:2006itc}
\bibitem{BaBar:2006itc}
B.~Aubert \textit{et al.} [BaBar],
Observation of a charmed baryon decaying to $D^0p$ at a mass near 2.94-GeV/c$^2$,
Phys. Rev. Lett. \textbf{98}, 012001 (2007).


%85\cite{Niu:2020gjw}
\bibitem{Niu:2020gjw}
P.~Y.~Niu, J.~M.~Richard, Q.~Wang and Q.~Zhao,
Hadronic weak decays of $\Lambda_c$ in the quark model,
Phys. Rev. D \textbf{102}, 073005 (2020).


%86\cite{Wang:2022zja}
\bibitem{Wang:2022zja}
K.~L.~Wang, Q.~F.~L{\"u}, J.~J.~Xie and X.~H.~Zhong,
Toward discovering the excited {\ensuremath{\Omega}} baryons through nonleptonic weak decays of $\Omega_c$,
Phys. Rev. D \textbf{107}, 034015 (2023).


%87\cite{Niu:2021qcc}
\bibitem{Niu:2021qcc}
P.~Y.~Niu, Q.~Wang and Q.~Zhao,
Study of heavy quark conserving weak decays in the quark model,
Phys. Lett. B \textbf{826}, 136916 (2022).


%88\cite{Niu:2025isf}
\bibitem{Niu:2025isf}
P.~Y.~Niu, Q.~Wang and Q.~Zhao,
Revisit the diquark of $\Lambda_c$ in the $\Lambda_c\to \Lambda K^+$ and $\Lambda_c\to \Sigma^0 K^+$ processes,
[arXiv:2507.04393 [hep-ph]].
%0 citations counted in INSPIRE as of 25 Nov 2025

%89\cite{Niu:2025lgt}
\bibitem{Niu:2025lgt}
P.~Y.~Niu, Q.~Wang and Q.~Zhao,
Cabibbo-favored hadronic weak decays of the $\Xi_c$ in the quark model,
Phys. Rev. D \textbf{111},  093004 (2025).


%90\cite{Wang:2024ozz}
\bibitem{Wang:2024ozz}
K.~L.~Wang, J.~Wang, Y.~K.~Hsiao and X.~H.~Zhong,
Excited $\Omega$ hyperon in charmful $\Omega_b$ weak decays,
Phys. Rev. D \textbf{111},  114028 (2025).


%91\cite{Cao:2023csx}
\bibitem{Cao:2023csx}
Y.~Cao, Y.~Cheng and Q.~Zhao,
Resolving the polarization puzzles in $D^0 \to VV$,
Phys. Rev. D \textbf{109}, 073002 (2024).


%92\cite{Pervin:2006ie}
\bibitem{Pervin:2006ie}
M.~Pervin, W.~Roberts and S.~Capstick,
Semileptonic decays of heavy omega baryons in a quark model,
Phys. Rev. C \textbf{74}, 025205 (2006).


%93\cite{Pervin:2005ve}
\bibitem{Pervin:2005ve}
M.~Pervin, W.~Roberts and S.~Capstick,
Semileptonic decays of heavy lambda baryons in a quark model,
Phys. Rev. C \textbf{72}, 035201 (2005).


%94\cite{Chua:2019yqh}
\bibitem{Chua:2019yqh}
C.~K.~Chua,
Color-allowed bottom baryon to $s$-wave and $p$-wave charmed baryon nonleptonic decays,
Phys. Rev. D \textbf{100},  034025 (2019).


%95\cite{Ke:2019smy}
\bibitem{Ke:2019smy}
H.~W.~Ke, N.~Hao and X.~Q.~Li,
Revisiting $\Lambda _{b}\rightarrow \Lambda _{c}$ and $\Sigma _{b}\rightarrow \Sigma _{c}$ weak decays in the light-front quark model,
Eur. Phys. J. C \textbf{79}, 540 (2019).


%96\cite{Zhu:2018jet}
\bibitem{Zhu:2018jet}
J.~Zhu, Z.~T.~Wei and H.~W.~Ke,
Semileptonic and nonleptonic weak decays of $\Lambda_b^0$,
Phys. Rev. D \textbf{99}, 054020 (2019).


%97\cite{Gutsche:2018utw}
\bibitem{Gutsche:2018utw}
T.~Gutsche, M.~A.~Ivanov, J.~G.~K{\"o}rner and V.~E.~Lyubovitskij,
Nonleptonic two-body decays of single heavy baryons  $\Lambda_Q$, $\Xi_Q$, and $\Omega_Q$ $(Q=b,c)$ induced by $W$ emission in the covariant confined quark model,
Phys. Rev. D \textbf{98}, 074011 (2018).


%98\cite{Mannel:1992ti}
\bibitem{Mannel:1992ti}
T.~Mannel and W.~Roberts,
Nonleptonic $\Lambda_b$ decays at colliders,
Z. Phys. C \textbf{59}, 179-182 (1993).


%99\cite{Giri:1997te}
\bibitem{Giri:1997te}
A.~K.~Giri, L.~Maharana and R.~Mohanta,
Two-body nonleptonic $\Lambda_b$ decays with $1/M_Q$ corrections,
Mod. Phys. Lett. A \textbf{13}, 23-32 (1998).


%100\cite{Cheng:1996cs}
\bibitem{Cheng:1996cs}
H.~Y.~Cheng,
Nonleptonic weak decays of bottom baryons,
Phys. Rev. D \textbf{56}, 2799-2811 (1997)
[erratum: Phys. Rev. D \textbf{99}, 079901 (2019)].


%101\cite{Fayyazuddin:1998ap}
\bibitem{Fayyazuddin:1998ap}
Fayyazuddin and Riazuddin,
Two-body nonleptonic $\Lambda_b$ decays in quark model with factorization ansatz,
Phys. Rev. D \textbf{58}, 014016 (1998).


%102\cite{Ivanov:1997ra}
\bibitem{Ivanov:1997ra}
M.~A.~Ivanov, J.~G.~Korner, V.~E.~Lyubovitskij and A.~G.~Rusetsky,
Exclusive nonleptonic decays of bottom and charm baryons in a relativistic three quark model: Evaluation of nonfactorizing diagrams,
Phys. Rev. D \textbf{57}, 5632-5652 (1998).


%103\cite{Mohanta:1998iu}
\bibitem{Mohanta:1998iu}
R.~Mohanta, A.~K.~Giri, M.~P.~Khanna, M.~Ishida, S.~Ishida and M.~Oda,
Hadronic weak decays of $\Lambda_b$ baryon in the covariant oscillator quark model,
Prog. Theor. Phys. \textbf{101}, 959-969 (1999).


%104\cite{Zhang:2022iun}
\bibitem{Zhang:2022iun}
C.~Q.~Zhang, J.~M.~Li, M.~K.~Jia and Z.~Rui,
Nonleptonic two-body decays of $\Lambda_b \to  \Lambda_c \pi$, $\Lambda_c K$ in the perturbative QCD approach,
Phys. Rev. D \textbf{105}, 073005 (2022).


%105\cite{Li:2022hcn}
\bibitem{Li:2022hcn}
Y.~S.~Li and X.~Liu,
Investigating the transition form factors of $\Lambda_b \to \Lambda_c(2625)$ and $\Xi_b \to \Xi_c(2815)$ and the corresponding weak decays with support from baryon spectroscopy,
Phys. Rev. D \textbf{107}, 033005 (2023).


%106\cite{Chua:2018lfa}
\bibitem{Chua:2018lfa}
C.~K.~Chua,
Color-allowed bottom baryon to charmed baryon nonleptonic decays,
Phys. Rev. D \textbf{99}, 014023 (2019).


%107\cite{Buchalla:1995vs}
\bibitem{Buchalla:1995vs}
G.~Buchalla, A.~J.~Buras and M.~E.~Lautenbacher,
Weak Decays beyond Leading Logarithms,
Rev. Mod. Phys. \textbf{68}, 1125-1144 (1996).


%108\cite{Mitroy:2013eom}
\bibitem{Mitroy:2013eom}
J.~Mitroy, S.~Bubin, W.~Horiuchi, Y.~Suzuki, L.~Adamowicz, W.~Cencek, K.~Szalewicz, J.~Komasa, D.~Blume and K.~Varga,
Theory and application of explicitly correlated Gaussians,
Rev. Mod. Phys. \textbf{85}, 693-749 (2013).


%109\cite{Varga:1997xga}
\bibitem{Varga:1997xga}
K.~Varga and Y.~Suzuki,
Solution of few body problems with the stochastic variational method: 1. Central forces,
Comput. Phys. Commun. \textbf{106}, 157-168 (1997).


%110\cite{Varga:1995dm}
\bibitem{Varga:1995dm}
K.~Varga and Y.~Suzuki,
Precise Solution of Few Body Problems with Stochastic Variational Method on Correlated Gaussian Basis,
Phys. Rev. C \textbf{52}, 2885-2905 (1995).


%111\cite{Zhong:2024mnt}
\bibitem{Zhong:2024mnt}
H.~H.~Zhong, M.~S.~Liu, R.~H.~Ni, M.~Y.~Chen, X.~H.~Zhong and Q.~Zhao,
Unified study of nucleon and {\ensuremath{\Delta}} baryon spectra and their strong decays with chiral dynamics,
Phys. Rev. D \textbf{110}, 116034 (2024).


%112\cite{Zhong:2025oti}
\bibitem{Zhong:2025oti}
H.~H.~Zhong, M.~S.~Liu, L.~Y.~Xiao, K.~L.~Wang, Qi-Li and X.~H.~Zhong,
$\Omega_c$ baryon spectrum and strong decays in a constituent quark model,
[arXiv:2502.13741 [hep-ph]].
%2 citations counted in INSPIRE as of 25 Nov 2025

%113\cite{Xiao:2013xi}
\bibitem{Xiao:2013xi}
L.~Y.~Xiao and X.~H.~Zhong,
$\Xi$ baryon strong decays in a chiral quark model,
Phys. Rev. D \textbf{87}, 094002 (2013).


%114\cite{Altmannshofer:2008dz}
\bibitem{Altmannshofer:2008dz}
W.~Altmannshofer, P.~Ball, A.~Bharucha, A.~J.~Buras, D.~M.~Straub and M.~Wick,
Symmetries and Asymmetries of $B \to K^{*} \mu^{+} \mu^{-}$ Decays in the Standard Model and Beyond,
JHEP \textbf{01}, 019 (2009).


%115\cite{Ni:2023lvx}
\bibitem{Ni:2023lvx}
R.~H.~Ni, J.~J.~Wu and X.~H.~Zhong,
Unified unquenched quark model for heavy-light mesons with chiral dynamics,
Phys. Rev. D \textbf{109},  116006 (2024).


%116\cite{Zhong:2008kd}
\bibitem{Zhong:2008kd}
X.~h.~Zhong and Q.~Zhao,
Strong decays of heavy-light mesons in a chiral quark model,
Phys. Rev. D \textbf{78}, 014029 (2008).


%117\cite{CDF:2011aa}
\bibitem{CDF:2011aa}
T.~Aaltonen \textit{et al.} [CDF],
Measurement of the branching fraction ${\mathcal{B}}(\Lambda^0_b\rightarrow \Lambda^+_c\pi^-\pi^+\pi^-)$ at CDF,
Phys. Rev. D \textbf{85}, 032003 (2012).


%118\cite{LHCb:2011poy}
\bibitem{LHCb:2011poy}
R.~Aaij \textit{et al.} [LHCb],
Measurements of the Branching fractions for $B_{(s)} \to D_{(s)}\pi\pi\pi$ and $\Lambda_b^0 \to \Lambda_c^+\pi\pi\pi$,
Phys. Rev. D \textbf{84}, 092001 (2011)
[erratum: Phys. Rev. D \textbf{85}, 039904 (2012)].


%119\cite{Oudichhya:2023awb}
\bibitem{Oudichhya:2023awb}
J.~Oudichhya and A.~K.~Rai,
Spin{\textendash}parity identification of newly observed singly charmed baryons in Regge phenomenology,
Eur. Phys. J. A \textbf{59}, 123 (2023).


%120\cite{Jakhad:2023mni}
\bibitem{Jakhad:2023mni}
P.~Jakhad, J.~Oudichhya, K.~Gandhi and A.~K.~Rai,
Identification of newly observed singly charmed baryons using the relativistic flux tube model,
Phys. Rev. D \textbf{108}, 014011 (2023).


%121\cite{Kim:2021ywp}
\bibitem{Kim:2021ywp}
Y.~Kim, Y.~R.~Liu, M.~Oka and K.~Suzuki,
Heavy baryon spectrum with chiral multiplets of scalar and vector diquarks,
Phys. Rev. D \textbf{104},  054012 (2021).


%122\cite{Shah:2016nxi}
\bibitem{Shah:2016nxi}
Z.~Shah, K.~Thakkar, A.~K.~Rai and P.~C.~Vinodkumar,
Mass spectra and Regge trajectories of $\Lambda_{c}^{+}$, $\Sigma_{c}^{0}$, $\Xi_{c}^{0}$ and $\Omega_{c}^{0}$ baryons,
Chin. Phys. C \textbf{40},  123102 (2016).


%123\cite{Belle:2022hnm}
\bibitem{Belle:2022hnm}
Y.~B.~Li \textit{et al.} [Belle],
Evidence of a New Excited Charmed Baryon Decaying to $\Sigma_c(2455)^{0,++} \pi^{\pm}$,
Phys. Rev. Lett. \textbf{130},  031901 (2023).


%124\cite{Luo:2025sns}
\bibitem{Luo:2025sns}
X.~Luo, S.~W.~Zhang, H.~X.~Chen, A.~Hosaka, N.~Su and H.~M.~Yang,
A short review on QCD sum rule studies of $P$-wave single heavy baryons,
[arXiv:2510.13013 [hep-ph]].
%1 citations counted in INSPIRE as of 25 Nov 2025



\end{thebibliography}

%\end{spacing}

\end{document}